\documentclass[final,5p]{elsarticle}

\usepackage{algorithm}
\usepackage{algpseudocode}
\usepackage{graphicx}
\usepackage{multirow}
\usepackage{url}
\usepackage{amsmath}
\usepackage{amssymb}
\usepackage{epstopdf}
\usepackage{amsthm}
\usepackage{color}
\usepackage{enumerate}

\fontfamily{pcr}\selectfont

\begin{document}

\begin{frontmatter}


\title{How Can Applications of Blockchain and Artificial Intelligence Improve Performance of Internet of Things? -- A Survey}


\author{Priyanka Bothra}
\address{Department of Information Technology,\\ Techno International New Town, Kolkata, INDIA 700156}
\ead{priyankabothra1502@gmail.com}

\author{Raja Karmakar\corref{mycorrespondingauthor}}
\address{Department of Electrical Engineering,\\ ETS, University of Quebec, Montreal, Canada}
\ead{raja.karmakar.1@ens.etsmtl.ca}

\author{Sanjukta Bhattacharya}
\address{Department of Information Technology,\\ Techno International New Town, Kolkata, INDIA 700156}
\ead{sbhattacharya.tict@gmail.com}

\author{Sayantani De}
\address{Department of Computer Science and Engineering,\\ Techno International New Town, Kolkata, INDIA 700156}
\ead{sayantani.de@tict.edu.in}

\cortext[mycorrespondingauthor]{Corresponding author}


\begin{abstract}

In the era of the Internet of Things (IoT), massive computing devices surrounding us operate and interact with each other to provide several significant services in industries, medical as well as in daily life activities at home, office, education sectors, and so on. The participating devices in an IoT network usually have resource constraints and the devices are prone to different cyber attacks, leading to the loopholes in the security and authentication. As a revolutionized and innovated technology, blockchain, that is applied in cryptocurrency, market prediction, etc., uses a distributed ledger that records transactions securely and efficiently. To utilize the great potential of blockchain, both industries and academia have paid a significant attention to integrate it with the IoT, as reported by several existing literature. On the other hand, Artificial Intelligence (AI) is able to embed intelligence in a system, and thus the AI can be integrated with IoT devices in order to automatically cope with different environments according to the demands. Furthermore, both blockchain and AI can be integrated with the IoT to design an automated secure and robust IoT model, as mentioned by numerous existing works. In this survey, we present a discussion on the IoT, blockchain, and AI, along with the descriptions of several research works that apply blockchain and AI in the IoT. In this direction, we point out strengths and limitations of the related existing researches. We also discuss different open challenges to exploit the full capacities of blockchain and AI in designing an IoT-based model. Therefore, the highlighted challenging issues can open the door for the development of future IoT models which will be intelligent and secure based on the integration of blockchain and AI with the IoT. 

\end{abstract}

\begin{keyword}
Internet of Things; Blockchain; Artificial Intelligence
\end{keyword}

\end{frontmatter}

\section{Introduction}
\label{sec:intro}

The revolution of incorporating billions of varieties of devices and digital information in real world~\cite{routh2018survey} has been brought by the Internet of Things (IoT). The IoT is a fast developing field in the history of computation, with an estimation of $50$ billion devices to be handled by the IoT at the end of the year of $2020$~\cite{al2020survey}. The `things' means physical devices, such as vehicles, televisions, watches, machines, etc., which are connected to the Internet. The IoT brings different devices under an umbrella, where different functionalities are provided by the devices with the help of inter-device communications. The things are connected to sensors to sense the physical environment and perform a communication with other things~\cite{hejazi2018survey}. The things (or devices) sense data to carry out any predefined action by monitoring the environment. Users access the devices with the help of the Internet and are notified by the execution of functions in the IoT; and accordingly, the users can take action for controlling the environment. Therefore, all the devices are connected together by using the Internet and collectively perform a function or more than one function to facilitate the use of collection of devices in order to fulfill an objective~\cite{dhanalaxmi2017survey}. The IoT can boost the growth of manufacturing and productivity in traditional companies (like transport, retail, etc.) and industries, by enhancing the efficiency of the architectures and processing mechanisms. The expansion of the IoT is booming in economics, technology, education, healthcare, and so on. This growth of the IoT leads to the revolution in the industry that is know as Industry 4.0~\cite{xu2018survey}.

Blockchain is an emerging Distributed Ledger Technology (DLT) following a decentralized architecture, that allows secure, immutable, and anonymous transactions~\cite{yu2018virtualization,belotti2019vademecum,gorkhali2020blockchain,lu2019blockchain,zhang2020review,nanayakkara2021methodology}. As reported in~\cite{gorkhali2020blockchain}, the research in blockchain requires a significant effort to develop new technologies and methodologies in order to integrate blockchain with other frameworks. Authors in~\cite{lu2019blockchain} review the studies on the blockchain, and highlight possible trends and challenges in the blockchain. The review~\cite{zhang2020review} provides a comprehensive overview of existing researches on blockchain along with the IoT, Industry 4.0, and Business Analytics. The methodology discussed in~\cite{nanayakkara2021methodology} provides a blockchain platform in the direction of solving enterprise issues.
According to National Institute of Standards and Technology (NIST), the blockchain technology provides a digital immutable ledger system that is implemented in distributed architectures (i.e., without any central repository in the model) and without involving a central authority~\cite{wu2019comprehensive}. Blockchain makes an effective use of distributed systems for different types of application domains, such as from a cryptocurrency system to an industrial system that requires decentralized, trusted, robust, and making automated decisions in different multi-stakeholder situations~\cite{belotti2019vademecum,wu2019comprehensive}. Blockchains help in issues created by third parties in financial transactions in the IoT, where the primary concerns are additional security risks and a lack of anonymity for users~\cite{ensor2018blockchains}. Networking is a fundamental technique that has a censorious role in designing and implementing blockchain, where blocks are chained with hash functions in order to impose privacy, security, and authentication in transactions.

Artificial Intelligence (AI) facilitates the development of intelligent systems by engaging Machine Learning (ML) techniques~\cite{al2020survey,lu2019artificial,zhang2021study}. An automated system reduces the intervention of human beings and takes decisions quickly based on the present situation. Thus, an intelligent system overcomes errors which can be introduced by human beings. The proper exploration and exploitation of the full capability of AI in designing future wireless technologies, such as 6th generation (6G), are an exceptionally hot research topic throughout the world~\cite{al2020survey}. AI ensures intelligent resource management through automatic adaptation capabilities and powerful learning~\cite{lin2020artificial}. The authors in~\cite{lu2019artificial} provide an extensive survey of AI considering industrial achievements and fundamental algorithms of AI. The overview of the scope of AI is presented in~\cite{zhang2021study} using different applications, drivers, and technologies. The work~\cite{chen2021explore} highlights different factors related to innovation characteristics of AI with external environment and organizational capabilities. The survey~\cite{hussain2020machine2} discusses the current solutions and research challenges in the direction of the application of ML for resource management in IoT and cellular networks.
In AI, machine learning techniques study the training dataset for future inferences by considering that the future data will have the same probability distribution or the identical feature space as the training dataset~\cite{shao2014transfer}. In ML, there are primarily three types of learning mechanisms -- unsupervised, supervised, and reinforcement learning. Each of these learning approaches has a specific field of application, and therefore the appropriate ML algorithm needs to be exploited based on the demand of the problem, in order to build an intelligent model for solving the problem.

The IoT economy demands a way for devices and sensors to perform monetary transactions for providing services without the involvement of third parties (e.g., Peer-to-Peer (P2P) and Machine-to-Machine (M2M))~\cite{miraz2015review}. Moreover, with the introduction of the fifth generation (5G), the exchanges and transactions in the IoT will primarily require an internet-based network, where any payment or billing system will need to be considered. In this direction, blockchain is an effective mechanism since it provides a distributed ledger technology for allowing immutable and anonymous transactions, by maintaining security, privacy, authentication, trustworthy, tamper-proof, etc~\cite{ensor2018blockchains,fan2020performance,belotti2019vademecum}. In addition, to store a massive volume of data generated by the devices in the IoT, the cloud computing technology can be used. Furthermore, to reduce the effective response time, the computation can be conducted in the edge in an IoT architecture~\cite{rafique2020complementing}. 
The work presented in~\cite{priyanka2020integrating} focuses on the development of an IoT-based intelligent framework to perform online control of the rate of pressure and flow in fluid transportation systems. The association of digital twin with the IoT can lead to the integration of the real world and digital world, which will play the key role in Industry 4.0~\cite{jiang2021digital}.
Due to the involvement of blockchain along with cloud or edge in the IoT, the design complexity is increased, where a huge volume of data needs to be handled efficiently by minimizing the losses and ensuring the security related constraints~\cite{gai2020blockchain,saad2020exploring,qiu2020networking}. Additionally, with the integration of billions of devices, the IoT has an objective to implement a communication between those devices with a minimal human intervention~\cite{al2020survey,diedrichs2018prediction}. Therefore, AI is an effective choice to make smart systems that can intelligently interact with each other in IoT paradigms by imposing smart access controls, where a massive volume of data will be handled with a minimum loss and maximum security. Hence, the performance of an IoT-based model can be enriched by exploiting both blockchain and AI, focusing on achieving smart mechanisms to deal with high data volume, adaptive communications and secured access controls~\cite{huang2019achieving,lin2019making,shen2019privacy,chai2020hierarchical,rahman2019blockchain,mohanta2020survey,wang2019video,samaniego2017internet,qiu2020networking,zhang2021study}. However, in order to enrich IoT systems in future, a thorough study of the impacts of using blockchain and AI on the performance of the IoT is required to identify the challenges in this domain. 

\textbf{Objective and Contributions of this Survey:} Since an IoT constructs the backbone for bringing and communicating several devices under a common architectural model, the overall performance of the IoT needs to be investigated after integrating it with blockchain and AI. There are several existing research works, such as~\cite{niu2019blockchain,khan2019journey,liu2020fabric,doku2019lightchain,guin2018ensuring,hao2018fastpay,le2019lightweight}, and so on, which apply blockchain in the IoT. The works like~\cite{diedrichs2018prediction,zhang2017machine,li2019system,xiao2018iot,li2018learning,hussain2020new} etc., make intelligent IoT models by using AI. Both blockchain and AI are exploited by several existing literature like~\cite{huang2019achieving,lin2019making,shen2019privacy,chai2020hierarchical,rahman2019blockchain,qiu2020networking} etc., in order to design secure, distributed, and smart IoT models. 
By considering such existing related research works, we present this survey to understand the performance improvements and issues associated with the integration of blockchain and AI with IoT systems. Specifically, the contributions of the survey are mentioned as follows:
\begin{itemize}
 \item We study existing related works that deal with the applications of blockchain and AI in the IoT. In order to highlight the performance improvement of the IoT due to the applications of blockchain and AI in the IoT, we present an extensive discussion on the contributions of existing works by categorizing the integration of the IoT with blockchain and AI in three directions:
 \begin{enumerate}
   \item Integration of IoT with blockchain
   \item Integration of IoT with AI, and
   \item Integration of IoT with blockchain and AI.
 \end{enumerate} 
 \item We highlight strengths and weaknesses of existing works which integrate the IoT with blockchain and AI. In this context, the discussion of such integration is presented concentrating on several important parameters such as decentralization, security, privacy, authentication, efficiency, robustness, etc.
 \item We discuss different research challenges and future research scopes in designing effective IoT frameworks integrating with the blockchain technology and AI, which can lead to efficient IoT models with a significant improvement of the performances of applications provided by the IoT.
\end{itemize}

\textbf{Uniqueness of this Survey:} To the best of our knowledge, this survey is the first in the direction of studying the individual impacts of blockchain and AI, and their collective impacts on the performance of the IoT. In addition, this survey identifies the research challenges and future research directions, considering the integration of the IoT with the blockchain technology and AI mechanisms. 

\textbf{Organization of this Survey:} The rest of this survey is organized as follows. In \S~\ref{sec:background}, we present descriptions on the IoT, blockchain, and AI. \S~\ref{sec:IoT_BC} discusses existing literature that apply blockchain in IoT frameworks. In \S~\ref{sec:IoT_AI}, we highlight existing research works that integrate IoT and AI. \S~\ref{sec:IoT_AI_BC} presents a detailed discussion of existing researches that consider the integration of both blockchain and AI with IoT architectures. In the applications of blockchain and AI in the IoT, different research challenges and future research directions are discussed in \S~\ref{sec:challenge}. Finally, we present the conclusion of the survey in \S~\ref{sec:conc}.

\section{IoT, Blockchain and AI: A Generic Description}
\label{sec:background}

In this section, we present descriptions of the IoT, blockchain, and AI, along with their fields of applications.

\subsection{Internet of Things}

The IoT is the latest technology evolving to connect heterogeneous devices for communication. The IoT is known as a growing technology through which anything can be communicated and connected to each other anywhere and anytime. The IoT defines an environment where computing and communication systems are embedded to each other to achieve specified goals by interchanging and collecting data. The IoT technology refers to a scenario where sensors, computation ability, and network are integrated with each other to consume, generate, and exchange data with a minor or without human interference.
The generation of advanced technological age is possible using the IoT where things in real world are enabled using various applications. The IoT is also able to provide a network-based environment where sharing of information, handling of resources and storage, properly organization of data, etc., can be done quickly and efficiently~\cite{neshenko2019demystifying}. This environment can be distributed in any of IoT-based applications areas.

\subsubsection{Different Applications in IoT}

\begin{figure}[!t]
	\centering
	\includegraphics[width=\linewidth]{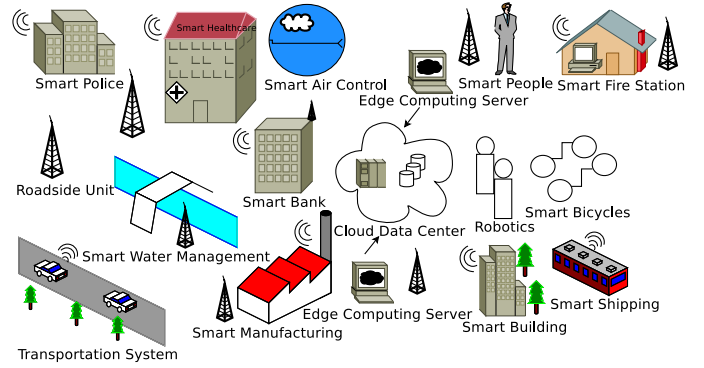}
	\caption{General applications of IoT}
	\label{fig:iot}
\end{figure}

The IoT provides a variety of applications in different segments of daily life. The bunch of applications is generally recognized as Cyber Physical System (CPS) which includes smart city, smart transportation, smart education, smart industry, smart healthcare, smart agriculture, smart security, etc., as shown in Fig.~\ref{fig:iot}. The major applications of the IoT are discussed as follows~\cite{hejazi2018survey}. 
\begin{itemize}
 \item Assisted Driving in one of the segment where advantages of the IoT are clearly visible. 
 \item In social networking segment, the location tracking in real time honestly provides various activities of users. 
 \item Another application is mobile ticketing where Near Field Communication (NFC) tag is coupled with mobile phones. Various transport related information in the NFC tag is retrieved automatically using mobile phones and transferred to users for satisfying their queries.                 
 \item In offices and homes, different sensors are deployed for various purposes such as temperature control or alarm systems, etc. 
 \item To save the valuable products from any types of losses and thefts, the precious products are tagged using the IoT such that, it can be identified and retrieved immediately after any of its displacement. 
 \item In the development of a smart city, each citizen can be facilitated using a smart infrastructure to safely enjoy a reliable, secure, and autonomous lifestyle.  
 \item Software Defined Internet of Things (SDIoT) is another advanced feather of the IoT, which is used to efficiently manage smart city infrastructure providing security, fault-tolerance, reliability, latency, versatile innovation, interoperability within different devices, circulation of traffic data on various devices, etc. 
 \item Intelligent and automated management is used to securely control road traffic. Different sensors and microcontrollers, such as Arduino microcontroller, proximity sensor, etc., are used for the detection of vehicles in a crowded area and also for the prediction of the excess time~\cite{dhanalaxmi2017survey}. 
The IoT introduces Vehicle Ad hoc Network (VANET) which combines the wireless technology and automobile infrastructure. 
 \item The application of the IoT in Healthcare sector is gaining popularity day by day. The continuous monitoring of the patients who are suffering from critical and rare diseases is absolutely necessary for their proper diagnosis. 
The IoT provides an optimum level of facility which is truly beneficial for both healthcare providers and patients. 
Emergency alerts or services are implemented using edge computing or fog computing, and Wireless Body Area Network (WBAN) in real time environments. 
\end{itemize}

\subsubsection{Architecture of IoT}

The architecture of the IoT focuses on some factors such as reliability, Quality of Service (QoS), integrity, and confidentiality. This architecture is also defined as a layered architecture known as perception, network, middleware, application, and business layers. All the service managements related to IoT applications are handled by the business layer. Different types of IoT oriented applications, such as smart city, are managed by the application layer. The middleware layer is responsible to store all the information which generally comes from lower layers. The function of the network layer is to securely store all the confidential information which comes from various sensors, and also propagate information to the upper layers. The lowest level layer which is defined as the perception layer incorporates with different sensor devices, such as ZigBee, Radio-Frequency Identification (RFID), etc., and the lowest layer is also responsible to collect the required information from the sensor devices. 

\subsubsection{Protocols in IoT}

\begin{figure*}[!t]
	\centering
	\includegraphics[width=0.85\linewidth]{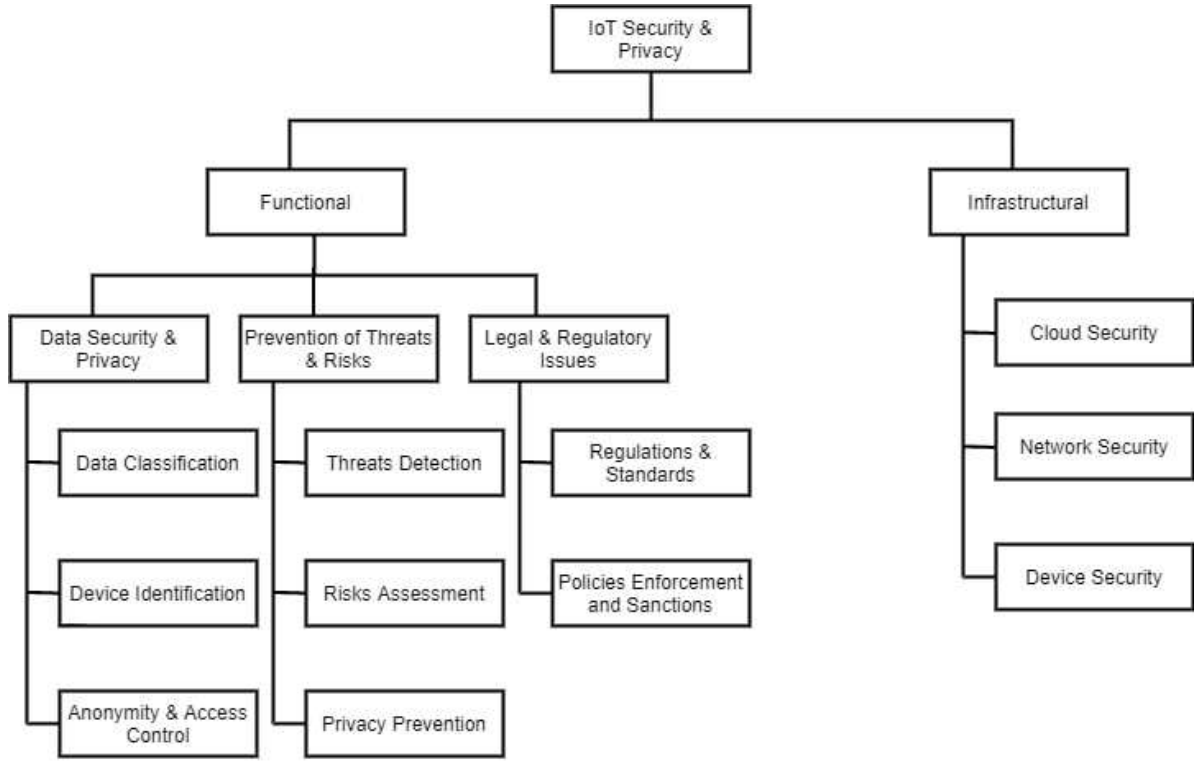}
	\caption{Taxonomy of IoT security and privacy}
	\label{fig:taxonomy}
\end{figure*}

Protocols are also crucial factors in the IoT-based infrastructure for the interaction between IoT gateways.
Some protocols are commonly used, such as Constrained Application Protocol (CoAP), which are used for constrained nodes and networks, and they are defined as web transfer protocols. Another protocol, Message Queuing Telemetry Transport (MQTT), which is known as client server messaging protocol, is also used to provide ordered lossless connections~\cite{datta2017survey}.

Another protocol, known as Advanced Message Queuing Protocol (AMQP), is responsible for message routing, message orientation, and message security mainly in the application layer. Extensible Messaging and Presence Protocol (XMPP) is also an important protocol which is accountable for various real time applications. Several security measures are associated with the IoT architecture and these security measures are achieved using several technologies, such as ZigBee, which provide the reliable communication, a minimum cost low power, etc. It is a two-way wireless personal area network (PAN) concept which is applicable for short range applications. 
Bluetooth is used for radio frequency communication for the short range. 
WiFi allows the internet in the shape of radio frequency. WiFi consists of several security modes which are responsible for filtering of protocols, assigning of IP addresses, WiFi-based security, filtering of MAC addresses, Shield Service Set Identifier (SSID), etc. 
Taxonomy of the IoT security and privacy is illustrated in Fig.~\ref{fig:taxonomy}.


\subsection{Blockchain}

Blockchain elevates the developments of Industrial world. It has designed and processed new business models which were impossible to shape few years ago. Blockchain secured many segments, like finance, healthcare, manufacturing, etc., through the effect of current business models. In organizations, the business, individuals, software agents, and a group of users are supported by many methods of blockchain for sharing secure records and transactions among them. Blockchain strengthens a group of individuals to make a settlement on a definite activity and this settled activity can also be registered, secured, shared, traced, tracked, and measured through blockchain. This agreed activity may be a payment deal, purchase activity, and medical lab-tests for a patient, voting activity, and contract agreement or it may be the supply chain of the process of planning, implementing, controlling, storage, services, and consumption of goods. As a technology, blockchain unites the benefits of both equal-level networks and the ordered steps of mathematical problems of the confidentiality of messages, integrity of messages, sender authentication, and assurances to make sure the reliability of the managed agreements. Agreed, approved, and recorded activity cannot be changed by any particular participant without connecting others. Blockchain can assure the state of error free, arranged functions over time. 
Connecting strongly two or more entities to register an agreement securely of definite tasks without knowing each other is one of the main characteristics of blockchain. This agreement cannot be recreated or removed by anyone in that group. But, they can only review the previous transactions easily without removing and modifying it. Activities can be easily recorded and preserved securely through blockchain by a group of people or organizations~\cite{al2019blockchain}. Moreover, the process of maintenance of blockchain makes an unchanged history of activities of a group's members. The process, known as 'mining', is used to authenticate and to grow dependency of the supervised agreements which are attached in the chain. Many industrial applications take support the internet transactions of blockchain. Blockchain is a developing technology which was previously created in Bitcoin area. But nowadays blockchain is applicable in multiple sectors by its secure and updated structure.

\begin{figure}[!t]
	\centering
	\includegraphics[width=\linewidth]{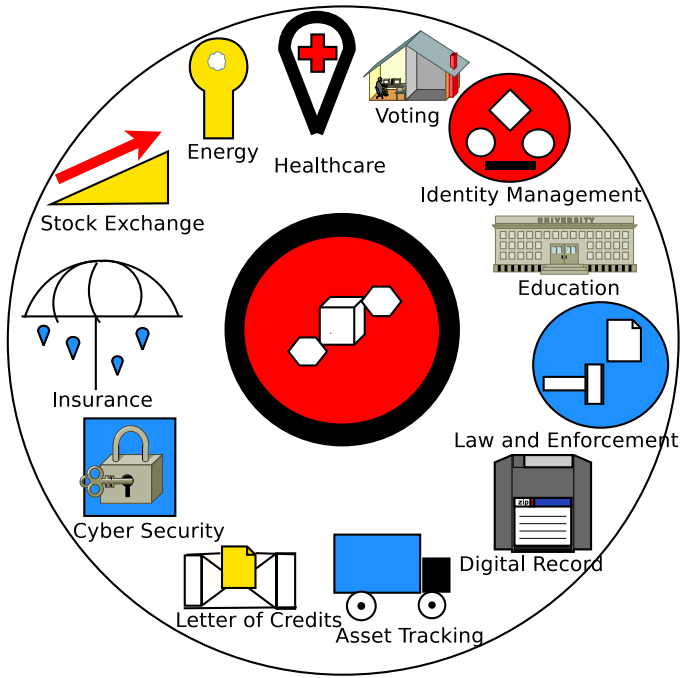}
	\caption{Generic applications of blockchain}
	\label{fig:blockchain}
\end{figure}

Blockchain technologies can be utilized in distinct sets of applications which are shown in Fig.~\ref{fig:blockchain}. Blockchain technology can find remedies in multiple sectors like voting, healthcare, supply chain, identity management, energy resources, governance, etc. 
Different applications of blockchain are as follows.

\begin{itemize}
 \item Blockchain can be served in healthcares by its ability to identify and measure of drugs and patients' data. In the pharmaceutical industry, inauthentication is one of the vital problem. By the Health Research Funding Organizations report, it was disclosed that $10\%$ to $30\%$ of the medicines have sold in the economically advanced countries. World Health Organization (WHO) evaluated that $16\%$ of drugs have improper mixture of substances and $17\%$ of drugs are comprised inexact level of necessary ingredients. So, these kind of medicines can be harmful to patients and will not cure them from diseases, rather they cause fatality. 
All these major issues can be directed towards workable solutions by blockchain technology because all the data and time of the agreements are recorded to the developed register, which can observe data and make information unchangeable. Blockchain can grow a better system to maintain the previous and present data of patients in a secure manner.
 \item The implementation of blockchain technology is associated with the energy industry through the major uses of microgrids. Blockchain can make easier to record and verify the power selling and buying deal in the small power supply grid containing a distributed generation of electricity. 
Moreover, energy related applications of blockchain have the capability to lower energy costs and to resume functionality~\cite{casino2019systematic}. 
 \item Through demoralization and automatic technology, blockchain could be most appropriate in the stock exchange. Blockchain can lower costs by demolishing source of intermediaries and accelerating transaction agreements. Furthermore, blockchain technology can offer possible uses in clearing and settlement of a transaction by easing repetitions of paper works in the business. Blockchain technology overcomes the need of third-party regulators in transactions. 
 \item Blockchain can solve the distrustfulness in the segment of voting by providing a developed record to assure the countability of voters. 
 \item Blockchain technology can reduce administration costs in the insurance sector by the arrangement, purchase, submission, claims, and record of insurance policies between clients, policy holders, and different companies. Furthermore, blockchain empowers the insurance industry to globally increase the productivity. 
 \item Blockchain technology can be an effective system for protecting the online identification. This technology can assure to save an individual's identity information from being theft and dishonest activities, by generating a updated technical platform. 
 \item The updated implementation of blockchain technology can able to speed up and secure the procedure of documentation in the trade finance in an assuring manner. 
\end{itemize}

In blockchain, the digital signature which is achieved using cryptography of private key is an important part to initiate a transaction. 
In blockchain networks, transactions are used to transfer digital assets within peers. These peers are responsible for choosing and validating the transactions. After the proper verification and validation procedure, one transaction is inserted in a block that is created by the miner node. The miner node has the ability to create a new block only when it can solve first a computational puzzle. After creating a new block in the network, it must be verified by all other peers and after that, it will be joined in existing chain. At this moment, transactions are specified as confirmed~\cite{monrat2019survey}. 

Two types of transactions are commonly used in blockchain: (i) Bitcoin transaction and (ii) Ethereum transaction. 
Bitcoin is a confidential, authentic, and integrated digital currency, which is performed over an accessible, public, and unidentified blockchain networks. In case of Ethereum transactions, it has sufficient fields to transfer ether (transferable amount) and messages which are beneficial to properly trigger the contracts.
Blockchain is the basic technology of a number of digital currency based on a cryptographic system. Chain of blocks can save information with digital indication in distributed and allocated networks. Dissolution, constancy, clarity, and the power of examining ability of blockchain characteristics build the transactions more protected and strongly unchangeable. Blockchain technology can be applied in various segments, such as healthcare, energy industry, stock market, voting, insurance, identity management, trade finance, social services, etc., to make feasible solutions in a secure manner~\cite{yang2019integrated}. Although blockchain-based remedies become popular for different business sectors, there are many implementations of blockchain, which need more analysis and improvement. Moreover, functions of blockchain need to remove the errors fast to develop more fruitful, effective, protected, and appealing form of blockchain technology in the long run. 

\subsection{Artificial Intelligence}

In the modern world, artificial intelligence is one of the prominent and rising technology, which can ease the tasks of human beings to complete the tasks efficiently and effortlessly. A person can use his intellect and creative mind through the performances of artificial intelligence which is the property of machines, computer programs, and systems. Artificial intelligence is often useful to find ways in problem solving and decision making independently. Learning, thinking, planning, and ability of taking actions independently are the fruitful outcomes of recent researches in AI. Some researches make conclusions that AI is wittier than human beings. On the contrary, some of the other researches criticize its functional ability and intelligence level in comparison with human brains. But large number of researches are involved to explore the possibilities of multiple capacities of AI in future. In recent times, AI is becoming one of the highest place of interests among the community of researchers~\cite{he2010survey}. Many successful researches in AI already revealed the various ways to apply it in solving problems easily associated with software engineering processes. By observing the rapid developments of AI in yesteryears, we all are hopeful for the positive revolutionary outcomes of researches to take digitalization on the top of the success history.

Artificial intelligence is introduced in various ways by many authors through multiple subjects and their scientific implementations in different times. `Thinking ability' is the common element among all the fields of AI. The term `intelligence' is actually the combination of the way of thinking, acquiring knowledge, and ability to implement in order to solve a given problem, as per requirements. AI achieves the capabilities and limitations of human beings in many tasks that were tough to be unable to accomplish in earlier times. Advancements in AI were achieved due to the increase of valid information, betterment in hardware, and the progress in the designs and operations of systems, which were planned precisely in ordered steps for the computational procedure. Nowadays, developers and researchers are allowed to code and test models fast by high quality open-source libraries which are the fruitful outcomes of advancements of AI~\cite{sheraz2020artificial}. Although some applications based on AI disappointed for their trust-related problems, developments in speech recognition, object detection, and many more have successfully led to their enlargement and filtration to the real-world applications, mostly in supervised learning by the rising performance of AI. 
Although the progress in the development of AI is increasing, there are lot of challenges that need to be overcomed to control the improvement in the applications of AI. 
Lot of questions and doubts have been arised repeatedly about the implementations and fruitful outcomes of AI because of distrustful results and unreliability in multiple segments like vision models, visual recognition, Natural Language Processing (NLP), etc. To protect future society, these hazards should be addressed fast~\cite{kadric2016accurate}. Although recent researches of deep learning have been performed noticeably, there are lot of darknesses in the area of improvements. There are various methods in AI, which are used by their applications in software engineering associated with software development processes, for solving problems. Although lot of researches already drew the conclusions that AI is notably difficult course to experiment through teaching, learning, and execution, the current improvements of AI can make the incompetent functions of many software products less severe. 

\begin{figure}[!t]
	\centering
	\includegraphics[width=\linewidth]{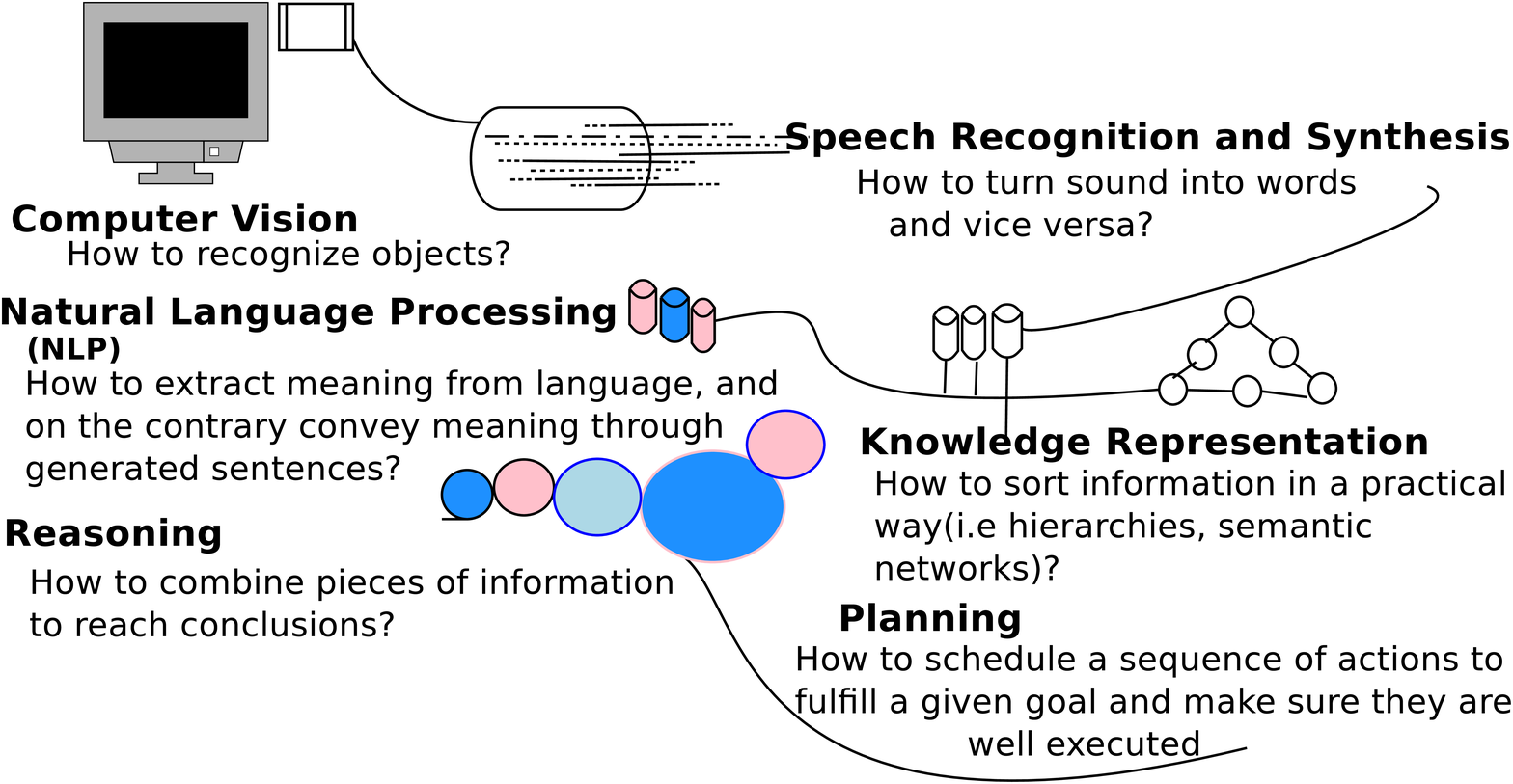}
	\caption{Working principle of AI}
	\label{fig:ai}
\end{figure}

Applications of AI are:
\begin{itemize}
 \item GPS navigations
 \item Customer services, finance, sales, and marketing
 \item Military applications
 \item Medicinal applications
 \item Space applications
 \item Industrial applications
 \item Telecommunication applications
\end{itemize}
An example of working principle of AI is shown in Fig.~\ref{fig:ai}.
Lot of researches revealed that AI can achieve remarkable discoveries in future~\cite{wang2020convergence}. Continuous improvements of AI will provide multiple options to make human life more easier. Although AI could not be $100\%$ reliable for human beings because of its executional errors in several times, AI is continuing its development consistently to provide accurate outcomes of its performances.

\section{Integration of IoT with Blockchain}
\label{sec:IoT_BC}

In this section, different uses of blockchain in IoT are presented in details. Tables~\ref{table:iotbc_dc},~\ref{table:iotbc_dc2},~\ref{table:iotbc_st},~\ref{table:iotbc_sec},~\ref{table:iotbc_sec2},~\ref{table:iotbc_priv},~\ref{table:iotbc_auth},~\ref{table:iotbc_tamperproof}, and~\ref{table:iotbc_auditable} highlight the strengths and weaknesses of the related research works which consider the integration of the IoT with blockchain in different perspectives such as decentralization, safety, security, privacy, authentication, tamper-proof, robustness, etc.

\subsection{Decentralization}

The IoT system along with blockchain is used in~\cite{ren2019using} to reduce accidents and improve traffic conditions by using decentralized peer-to-peer communication instead of cloud-based centralized system, so that data can be transmitted efficiently. Confidentiality, integrity, and availability of data are ensured by using blockchain which also makes external attacks negligible since users communicate and self-execute automatically. However, there are some vulnerabilities including power consumption and excessive overload of the network throughput, which are not taken into consideration in~\cite{ren2019using}.
In~\cite{nguyen2018blockchain}, it is observed that decentralized ledger technology is based on various cryptocurrencies such as bitcoin and Ethereum. Cryptocurrency provides a reliable, stable, and robust way to store, transmit, and validate data to ensure democracy, transparency, and safety in various organizations. 
With the development of the IoT 4.0, the impact of cryptocurrency would be enormous, and it will be applied in various fields like banking, healthcare, financial, etc. 
Using blockchain as the underlying communication architecture, a security auditing system is constructed in~\cite{cha2018iso}, which can predict and detect any anomalies to guarantee confidentiality of data, issue alerts for security incidents, collect evidences for auditing purposes, and probe event trajectory via data analysis in order to track the involved devices. It uses decentralized system to prevent tampering of data and help in encryption and decryption of data records. Using smart contract scheme on a single blockchain node, the effectiveness and efficiency of the security records are examined. 

Two Local Energy Market (LEM) mechanisms are proposed in~\cite{moniruzzaman2019blockchain} with artificial prosumers and these mechanisms are implemented on private blockchain in a decentralized manner based on economic evaluation. It promotes increase in the production of renewable energy and decrease in consumption to reduce usage spikes at peak-hour. Through experiments, it is found that prosumers energy behavior can affect their economic benefits; and if users' energy supply is increased and demand is decreased during peak-hour, their economic benefits increase, which helps in boosting the local economy. 
The decentralized WEB revolutionized the technologies such as blockchain by introducing symbiotic webs as discussed in~\cite{khan2019journey}. It helps in linking web pages with AI, NLP etc., and makes it transparent and efficient. 
Due to several characteristics of the IoT such as mobility, distributed deployment, and limited performance, centralized access control method does not support access control in a large scale environment. So, based on attributed-based access control (ABAC) and Hyperledger Fabric Blockchain framework, an access control system is proposed in~\cite{liu2020fabric} making use of blockchain's decentralization, traceability as well as tamper-proof property. The proposed system has a device authority model which ensures the security of the device using three smart contracts, and helps in maintaining high throughput along with ensuring data consistency.
In a network, to address the mining process which has limited resources, a blockchain-based solution is presented in~\cite{doku2019lightchain}, which verifies a transaction. To overcome the problem of Proof of Work (PoW) which arises due to limited resources as well as its expensiveness, a decentralized blockchain system is used, where the nodes of the IoT devices are grouped together as a cluster in the network, and multiple transactions occurring at the same time are verified making it more efficient. First Fit Decreasing (FFD) method is modified so that node which leaves the cluster can be replaced with proper fit and helps in overcoming the limitations of PoW.

A transaction settlement system, NormaChain, is proposed in~\cite{liu2018normachain}, for IoT-based e-commerce based on blockchain in order to provide security against collusion and cryptoanalysis, and ensures privacy. NormaChain overcomes various problems such as huge computational overhead and non-supervisability. It increases system scalability and transaction efficiency. A decentralized public key searchable encryption scheme (Decentralized PEKS scheme) for cryptosystems is designed, which helps in achieving crime traceability and illegal transactions. 
A Decentralized IoT Collectability Data Marketplace (DCDM) market model is presented in~\cite{nguyen2019enabling}, where various operational factors are involved such as contextual, operationality, geographical location, and data provider availability. It uses a combination of crowdsensing and blockchain. It employs crowdsensing to build a business model for trading the IoT and collect data on-demand from customers, whereas blockchain is used to ensure the privacy and transparency. 
Using community-based decentralized architecture, security is maintained and there is a reward mechanism to boost the performance. 

A decentralized application (DApp), as presented in~\cite{papadodimas2018implementation}, is used for sharing IoT data based on blockchain technology using smart contracts without the involvement of a third party. It is accessible to everyone. It applies Sensing-as-a-Service (SaaS) which provides opportunity for monetization of data by helping in selling sensor data easily. By using cryptocurrency, DApp ensures security and flexibility to the users transacting on peer-to-peer level. However, the system faces various issues, such as scaling and high delay in transaction confirmation, which are to be addressed.
BlendCAC is used in large scale IoT devices for access control processes to provide fine-grained, scalable, and lightweight control mechanisms. In~\cite{xu2018blendcac}, IoT devices control their own resources. Local private blockchain network is used and smart contracts are implemented for registration, revocation of the access authorization, and propagation. A capability delegation mechanism is proposed, which helps in access permission, and a token management strategy is also implemented. It is found that the system helps in validation and access control authorization in a decentralized and trustless IoT network.

\begin{table}[!t]
\caption{Contributions of Existing Works Integrating IoT and Blockchain, and Concentrating on Decentralization}
\centering
\begin{tabular}{|p{0.7cm}|p{7.1cm}|}
\hline
\textbf{Item} & \textbf{Contributions} \\
\hline
\cite{ren2019using} & Reduces accidents and improves traffic conditions by using decentralized peer-to-peer communications.\\
\hline
\cite{nguyen2018blockchain} & Cryptocurrency provides a reliable, stable, and robust way to store, transmit, and validate data.\\
\hline
\cite{cha2018iso} & Predicts and detects any anomalies to guarantee confidentiality of data.\\
\hline
\cite{moniruzzaman2019blockchain} & Uses a private blockchain and increases the production of renewable energy.\\
\hline
\cite{khan2019journey} & The decentralized WEB revolutionizes the technologies such as blockchain by introducing symbiotic webs.\\
\hline
\cite{liu2020fabric} & Ensures security of the device accessed, and helps in maintaining high throughput along with ensuring data consistency.\\
\hline
\cite{doku2019lightchain} & IoT devices are grouped together as a cluster, and multiple transactions are verified at the same time.\\
\hline
\cite{liu2018normachain} & Blockchain-based e-commerce provides security against collusion and cryptoanalysis, and ensures privacy.\\
\hline
\cite{nguyen2019enabling} & Employs blockchain-crowdsensing to build a business model for trading IoT and collect data on-demand from customers.\\
\hline
\cite{papadodimas2018implementation} & Shares IoT data based on blockchain technology using smart contracts without the involvement of third party.\\
\hline
\cite{xu2018blendcac} & Local private blockchain and smart contracts are implemented for registration, revocation of the access authorization and propagation.\\
\hline
\cite{ozyilmaz2019designing} & Based on the computing and storage capabilities, all IoT devices are integrated, which will lead to data-centric models.\\
\hline
\cite{choudhary2020convergence} & Potential business applications are explored in order to form a smart world.\\
\hline
\cite{xiong2020best} & An architecture is presented by combining IoT systems with blockchain for managing IoT data for further utilization like retrieving and auditing.\\
\hline
\end{tabular} 
\label{table:t1}
\end{table}

With the increased number of IoT devices, there is a need to create an IoT back-end, and to improve the method of communication for IoT gateways, which can be achieved using trustless and decentralized nature of blockchain providing peer-to-peer network, as discussed in~\cite{ozyilmaz2019designing}. 
Based on the computing and storage capabilities, all the IoT devices can be integrated, which will lead to data-centric models where data processing and application development are conducted by smart contracts.
In order to create a smart environment, various benefits, drawbacks as well as trending research topics of using the decentralized nature of blockchain along with IoT are explored in~\cite{choudhary2020convergence}. 
Potential business applications are explored in order to form a smart world. 
Conventional IoT systems suffered from various issues like lack of transparency, limited scalability, and single point of failure. In~\cite{xiong2020best}, a general architecture is presented by combining IoT systems with blockchain technologies to implement the decentralized nature of blockchain for managing IoT data for further utilization like retrieving and auditing. A case study of a resource allocation learning-based method is also proposed for intelligently managing data. 
Table~\ref{table:t1} highlights existing works which consider the integration of the IoT and blockchain concentrating on decentralization.

\subsection{Safety and Transparency}

The application of blockchain in the perspective of safety and transparency is highlighted in~\cite{nguyen2018blockchain}. Blockchain helps in digitally transferring money safely to the intended recipient by protecting against various external attacks, losses, and errors. With the development of the IoT 4.0, its impact would be enormous and it will be applied in various fields like banking, healthcare, financial, etc. 
The WEB revolutionized the technologies such as blockchain by introducing symbiotic webs making it transparent and efficient. The Industry 4.0 encompasses smart contracts, smart supply chain, smart products, and smart factory, as discussed in~\cite{khan2019journey}. Blockchain has various features, such as immutability and transparency, which make it distinguishable.
There are various blockchain-based IoT projects but there is no method to compare their performances based on energy consumption. Blockchain technologies provide trust while transacting at the cost of energy instead of manpower. A research to model the energy consumption behavior is done in~\cite{cole2018modeling} by collecting real-world data. The model reflects on various algorithms such as PoW, Proof of Stake (PoS), Stellar Consensus Protocol (SCP), and Ripple Protocol Consensus Algorithm (RPCA) based on linear models.

To improve the enactment of the IoT and its application in an uncontrolled, hostile environment, and to automate living conditions ensuring the privacy and security, a blockchain model is proposed in~\cite{muzammal2018study}. It provides various advantages to the IoT system, such as security, cost reduction, immutability, anonymity, robustness, creditability, trust building, authentication, verification, etc. However, in the proposed model, there are some challenges such as consensus mechanism and computational cost for verification of transactions.
A comprehensive analysis is presented in~\cite{pervez2019blockchain} in the direction of the disruptive innovation occurring in the field of smart contracts and blockchain. A shared distributed ledger ensures secure and tamper-proof transactions between various users in the network without involving third parties. With the application of Directed Acyclic Graph (DAG) and its multi-forked structure, high scalability, efficient provenance, and optimized validation are observed. 
Together the IoT and blockchain help in real-time monitoring of resources and ensure non-repudiation, privacy, accountability, predictive analysis, transparency, and stakeholder visibility.

\begin{table}[!t]
\caption{Contributions of Existing Works Integrating IoT and Blockchain, and Concentrating on Safety and Transparency}
\centering
\begin{tabular}{|p{0.7cm}|p{7.1cm}|}
\hline
\textbf{Item} & \textbf{Contributions} \\
\hline
\cite{nguyen2018blockchain} & With the development of IoT 4.0, its impact would be enormous and it will be applied in various fields like banking, healthcare, financial, etc.\\
\hline
\cite{khan2019journey} & Industry 4.0 encompasses smart contracts, smart supply chain, smart products, and smart factory.\\
\hline
\cite{cole2018modeling} & A research to model the energy consumption behavior is done by collecting real-world data.\\
\hline
\cite{muzammal2018study} & Improves the enactment of IoT and its application in an uncontrolled, hostile environment and automates living conditions ensuring privacy and security.\\
\hline
\cite{pervez2019blockchain} & A comprehensive analysis is presented in the direction of the disruptive innovation occurring in the field of smart contracts and blockchain.\\
\hline
\cite{vafiadis2019differentiating} & Increases the efficiency by improving the transparency of data.\\
\hline
\cite{li2018blockchain} & Converging cloud computing, IoT, blockchain, and data analytics, a prototype of transportation scheme is proposed.\\
\hline
\cite{novo2018blockchain} & Provides scalable, generic, and manageable access control system by implementing proof-of-concept.\\
\hline
\cite{cole2018modeling} & Uses logistics planner and smart contracts to secure trust, provide visibility, and help in condition-monitoring of assets.\\
\hline
\end{tabular} 
\label{table:t2}
\end{table}

Blockchain technologies implemented on the IoT system in~\cite{vafiadis2019differentiating} increase the efficiency by improving the transparency of data. It is found that cryptocurrency IOTA is not the best solution with respect to quality improvements. It ensures high scalability by storing a large amount of data but due to the usage of central server, it is prune to attacks. A study is done to recommend which platform is to used based on the demand.
Converging cloud computing, IoT, blockchain, and data analytics, a prototype of transportation scheme is proposed in~\cite{li2018blockchain}, where both private and public blockchains are used. Hyperledger fabric saves the massive trip data and Ethereum framework is implemented as an insurance and payment model. Based on a driver's behavior, insurance premium is accessed and deducted from IoT data, which is collected from mobile sensors. The proposed model helps in encouraging drivers to drive safely. It is evaluated after experiments that resource utilization is correlated with request load. The system is capable of handling high request load and helps in achieving better throughput.

To ensure scalability to constrained IoT devices, blockchain is used as it offers various strengths, such as transparency, auditability, decentralized consensus, security, etc., which make it an ideal component of IoT solutions. The work~\cite{novo2018blockchain} provides a scalable, generic, and manageable access control system. It also implements proof-of-concept (PoC) which enables various devices to connect to the same network. Due to the versatility of management hub nodes, high flexibility of the network is ensured but the network suffers from overhead of waiting due to the time required to issue control information by the blockchain network.
Smart logistics solution in~\cite{cole2018modeling} uses logistics planner and smart contracts to secure trust, provide visibility, and help in condition-monitoring of assets using distributed ML framework in supply chain management areas. It demonstrates liability, transparency, accountability, optimal planning, and traceability, for handling assets by different parties involved in the logistics of supply chain management with the help of various technologies such as IoT, blockchain, ML, and Big Data.
Table~\ref{table:t2} highlights existing works which consider the integration of the IoT and blockchain concentrating on safety and transparency.

\begin{table*}
\caption{Integration of IoT with Blockchain Concentrating on Decentralization}
\centering
\begin{tabular}{|p{2.8cm}|p{8.5cm}|p{4.5cm}|}
\hline
\textbf{Item} & \textbf{Strength} & \textbf{Weakness} \\
\hline

Qilei \textit{et al.} \cite{ren2019using} & Reduction of accidents and improve traffic conditions by using decentralized peer-to-peer communication instead of cloud based centralized system so that data can be transmitted efficiently & Vulnerabilities include -- (1) power consumption, and (2) excessive overload of network throughput\\
\hline

Nguyen \textit{et al.} \cite{nguyen2018blockchain} & (1) Highlight of the decentralized ledger technology that is behind various cryptocurrencies like bitcoin and Ethereum, (2) Discussion on digitally transferring money safely to the intended recipient by protecting against various external attacks, losses or errors & No path is discussed to the design of blockchain-based IoT 4.0\\ 
\hline

Shi-Cho \textit{et al.} \cite{cha2018iso} & (1) Using blockchain as the underlying communication architecture, a security auditing system is constructed, (2) Predicts and detects any anomalies to guarantee confidentiality
of data, issue alerts for security incidents & Lack of analysis of the proposed architecture on multiple blockchain node\\
\hline

Moniruzzaman \textit{et al.} \cite{moniruzzaman2019blockchain} &
(1) Two LEM mechanisms are proposed with artificial prosumers, (2) Implemented on private
blockchain in a decentralized manner based on economic evaluation,
(3) Promotes increase in the production of renewable energy and decrease in consumption to reduce usage spikes at peak-hour & Global energy consumption issues are not addressed\\
\hline

Khan \textit{et al.} \cite{khan2019journey} & (1) Highlights decentralized WEB revolution with blockchain by introducing symbiotic webs, (2) A path to Industry 4.0 that encompasses smart contracts, smart supply chain, smart products and smart factory & (1) Details of security issues are not explored, (2) Implementation details and challenges are not highlighted, (3) Storage management issues are not discussed\\
\hline

Han \textit{et al.} \cite{liu2020fabric} & (1) An access control system based on attributed-based access control (ABAC) and Hyperledger Fabric Blockchain framework by using of blockchain's decentralization, traceability and tamper-proof property, (2) A device authority model which ensures security of the device accessed, using smart contracts, (3) Helps in maintaining high throughput ensuring data consistency & (1) Storage constraint and scalability issues are not in details, (2) Power management is not addressed\\
\hline

Ronald \textit{et al.} \cite{doku2019lightchain} & (1) Addresses the mining process which have limited resources, (2) Blockchain based solution is shown verify a transaction, (3) Uses a decentralized blockchain system to overcome the problem of PoW which arises due to limited resources and its expensiveness, (4) FFD method is modified so that the node which leaves the cluster can be replaced with proper fit and can help in overcoming the limitations of PoW & (1) Lack of discussion on the improvement of consensus mechanism in IoT platform, (2) Storage management\\
\hline

Xiao \textit{et al.} \cite{liu2018normachain} & (1) Security against collusion and cryptoanalysis and ensures privacy, (2) Overcomes huge computational overhead and non-supervisability, (3) Increases system scalability and transaction efficiency, (4) Helps in achieving crime traceability and eliminating criminal and illegal transactions & Integrity and scalability issues are not in details, (2) Storage management\\
\hline

Duc-Duy \textit{et al.} \cite{nguyen2019enabling} & (1) DCDM market model is proposed, (2) Uses a combination of crowdsensing and blockchain, (3) Security is maintained, (4) A reward mechanism to boost the performance & (1) Time complexity due to the use of crowdsensing, (2) Storage access mechanism is not highlighted\\
\hline

\end{tabular} 
\label{table:iotbc_dc}
\end{table*}

\begin{table*}
\caption{Integration of IoT with Blockchain Concentrating on Decentralization (contd.)}
\centering
\begin{tabular}{|p{2.8cm}|p{8.5cm}|p{4.5cm}|}
\hline
\textbf{Item} & \textbf{Strength} & \textbf{Weakness} \\
\hline

Georgios \textit{et al.} \cite{papadodimas2018implementation} & (1) Uses decentralized application for sharing IoT data based on blockchain technology using smart contracts without the involvement of third party, (2) Ensures security and flexibility to the users transacting on peer-to-peer level using cryptocurrencies & (1) Scaling issues, (2) High delay in transaction confirmation\\
\hline

Ronghua \textit{et al.} \cite{xu2018blendcac} & (1) Targets a large scale IoT devices for access control processes to information, devices and services, (2) Provides fine-grained, scalable and lightweight control mechanism, (3) Local private blockchain network and smart contracts for registration, revocation of the access authorization and propagation, (3) A capability delegation mechanism to help in accessing permission, (4) Token management
strategy & (1) Transaction throughput management, (2) Storage management details, (3) Time bound analysis\\
\hline

Ozyilmaz \textit{et al.} \cite{ozyilmaz2019designing} & (1) Improves the method of communication for IoT gateways by using trust less and decentralized nature of blockchain providing peer-to-peer network, (2) Distributed storage service & (1) Scalability and reliability issues, (2) Transaction management, (3) Transaction throughput analysis\\
\hline

Choudhary \textit{et al.} \cite{choudhary2020convergence} & (1) Various benefits, drawbacks as well as trending research topics of using the decentralized nature of blockchain in IoT, (2) Potential business applications in the future directions, (3) Security features of blockchain in IoT & (1) Storage management issues, (2) Implementation challenges of blockchain in IoT\\
\hline

Zhang \textit{et al.} \cite{xiong2020best} & (1) A general architecture combining IoT systems with blockchain technologies to implement decentralized architecture, (2) A case study of a resource allocation learning based method for intelligently managing data & Analyzing the robustness of the architecture in different network scenarios\\
\hline
\end{tabular} 
\label{table:iotbc_dc2}
\end{table*}

\begin{table*}
\caption{Integration of IoT with Blockchain Concentrating on Safety and Transparency}
\centering
\begin{tabular}{|p{2.8cm}|p{8.5cm}|p{4.5cm}|}
\hline
\textbf{Item} & \textbf{Strength} & \textbf{Weakness} \\
\hline

Nguyen \textit{et al.} \cite{nguyen2018blockchain} & (1) Highlights reliability, stability and robustness to store, transmit and validate data to help various organizations ensuring democracy, transparency and safety, (2) Helps in digitally transferring money safely to the intended recipient by protecting against various external attacks and losses & (1) Implementation details of integrating blockchain with IoT are not explored, (2) Lack of responsiveness analysis\\
\hline 

Ryan \textit{et al.} \cite{cole2018modeling} & A research to model the energy consumption behavior by collecting real-world data which reflects on PoW, PoS etc. & (1) Scalability analysis, (2) Transaction throughput management\\
\hline 

Muzammal \textit{et al.} \cite{muzammal2018study} & Integrating blockchain with IoT provides various advantages to the IoT system such as security, cost reduction, immutability, anonymity, robustness, creditability, trust building, authentication, verification etc. & (1) Lack of details of consensus mechanism, (2) Computational cost for verification of transaction\\
\hline

Huma \textit{et al.} \cite{pervez2019blockchain} & Comprehensive analysis of the disruptive innovation occurring in the field of blockchain and IoT & Details of privacy, security, authentication, storage management and transaction throughput are not explored\\
\hline

Vafiadis \textit{et al.} \cite{vafiadis2019differentiating} & (1) Blockchain technologies implemented on IoT system increase the efficiency by improving the transparency of data, (2) Ensures high scalability by storing large amount of data but due to the usage of central server it is prune to attacks, (3) A study to recommend which platform to use based on the demand & (1) Storage constraint analysis considering practical scenario, (2) Response time and transaction throughput management\\
\hline

Zengxiang \textit{et al.} \cite{li2018blockchain} & (1) A prototype of transportation scheme by converging cloud computing, IoT, blockchain and data analytics, (2) Both private and public blockchain, (3) Resource utilization is correlated with request load, (4) The system is capable of handling high request load, (5) Helps in achieving better throughput & 
(1) Lack of details of consensus mechanism, (2) Storage management\\
\hline

Oscar \textit{et al.} \cite{novo2018blockchain} & (1) Provides scalable, generic, manageable access control system, (2) Implements PoC which enables various devices to connect to the same network, (3) High flexibility of the network is ensured & The network suffers from overhead of waiting due to the time required to issue control information by the blockchain network\\
\hline

Ryan \textit{et al.} \cite{cole2018modeling} & Smart logistics solution for smart contracts to secure trust, provide visibility and help in condition-monitoring of assets, (2) Demonstrates liability, transparency, accountability, optimal planning and traceability, for handling different parties involved in the logistics of supply chain management & (1) Details of transaction delay and throughput are not present, (2) Storage management\\
\hline

\end{tabular} 
\label{table:iotbc_st}
\end{table*}


\begin{table*}
\caption{Integration of IoT with Blockchain Concentrating on Security}
\centering
\begin{tabular}{|p{2.8cm}|p{8.5cm}|p{4.5cm}|}
\hline
\textbf{Item} & \textbf{Strength} & \textbf{Weakness} \\
\hline

Lei \textit{et al.} \cite{lei2019next} & (1) Provides a scalable, robust and secure network environment, (2) Useful for smart transportation-related application & (1) Implementation details with respect to storage utilization and responsiveness analysis, (2) Computational cost for verification of transaction \\
\hline 

Cha \textit{et al.} \cite{cha2018iso} & (1) Predicts and detects any anomalies to guarantee confidentiality of data, (2) Issues alerts for security incidents, (3) collect evidences for auditing purposes & The communication details in the proposed decentralized architecture \\
\hline

Niu \textit{et al.} \cite{niu2019blockchain} & (1) Efficient outsourcing services, (2) Efficient in computational costs as well as communication overheads & Storage management and responsiveness are not highlighted\\
\hline

Haseeb \textit{et al.} \cite{haseeb2019intrusion} & (1) Secure routing to improve data reliability against various malicious attacks & (1) Implementation details of the proposed routing mechanism, (2) Storage management\\
\hline

Srivastava \textit{et al.} \cite{srivastava2019light} & (1) Advance security and privacy to the remote patient monitoring, (2) Data storing over cloud by using advanced cryptographic techniques, (3) Double encryption scheme & (1) Scalability and integrity analysis, (2) Latency analysis in details\\
\hline

Agrawal \textit{et al.} \cite{agrawal2018continuous} & Prediction models LSTM is used to provide continuous security with seamless user authentication & Scalability and storage management analysis\\
\hline

Faika \textit{et al.} \cite{faika2019blockchain} & (1) Energy efficiency, (2) Provides key management, access control and reduces latency & Latency of the proposed scheme can be further reduced if instead of using cloud-based blockchain server, local blockchain server is used\\
\hline

Hao \textit{et al.} \cite{hao2018fastpay} & Protects the system from double-spending attacks & The implementation details with storage analysis\\
\hline

Huang \textit{et al.} \cite{huang2019b} & (1) Decreases power consumption, (2) Improves security as well as
transaction efficiency, (3) Data authority management system is designed to protect the confidentiality of data & Quality control of sensor data as well as storing large amounts of data\\
\hline

Mokhtar \textit{et al.} \cite{mokhtar2019blockchain} & (1) Provides a time-sensitive distributed system using
smart contracts by considering performance, time constraints and autonomy requirements, (2) Adapting power-efficient mechanism, (3) Potential to enable scalability and security for IoT use cases, (4) Provides flexibility to the distributed system, (5) Latency minimization & (1) Lack of storage management analysis, (2) Implementation constraints\\
\hline

Nguyen \textit{et al.} \cite{nguyen2019enabling} & (1) Boosts performance reward mechanism, (2) Payment can be made in an auditable and secure manner & Lack of scalability and integrity analysis\\
\hline

Sanju \textit{et al.} \cite{sanju2018energy} & Various blockchain platforms such as Ethereum and Hyperledger fabric are compared based on their consumption of energy for real workloads & Reliability, scalability, storage management etc. are not analyzed\\
\hline 

Urmila \textit{et al.} \cite{urmila2019comparitive} & (1) Increases the security and robustness by utilizing cryptographic algorithm, (2) Trustworthy and peer-to-peer platform to manage data efficiently without the involvement of third party, (3) Provides a common platform so that different IoT devices can communicate in a distributed manner securely and protect against malicious attack, (4) Ensures feasibility, fault tolerance and
scalability & Analysis of responsiveness considering different types of attacks\\
\hline 

\end{tabular} 
\label{table:iotbc_sec}
\end{table*}

\begin{table*}
\caption{Integration of IoT with Blockchain Concentrating on Security (contd.)}
\centering
\begin{tabular}{|p{2.8cm}|p{8.5cm}|p{4.5cm}|}
\hline
\textbf{Item} & \textbf{Strength} & \textbf{Weakness} \\
\hline

Dinesh \textit{et al.} \cite{dinesh2019conforming} & Communication constraints faced by blockchain is resolved by 5G communication & (1) Limited storage capacity of IoT system, (2) Poor scalability, (3) Time consumption while mining, (4) Increase of traffic overhead\\
\hline 

Choo \textit{et al.} \cite{choo2020blockchain} & (1) Helps in identity verification while maintaining trust, (2) Transmits messages securely without involving centralized third-party & Challenges in implementation such as security and computational costs\\
\hline


Lee \textit{et al.} \cite{lee2018towards} & (1) Secures the system from external attacks, (2) Preserves privacy and uses smart contracts in various IoT related applications & (1) Implementation details are not highlighted, (2) Lack of analysis of storage management issues and responsiveness, (3) Overhead analysis\\
\hline

Rashid \textit{et al.} \cite{rashid2019security} & (1) Optimization of the lifetime of the network by self-clustering methodology, (2) Helps in authorization, authentication and verification, (3) Improves safety and explores
upper layer communications, (4) Due to multi-layer approach the, network load, delay and computational load are reduced, (5) Communication efficiency is increased along with integrity and security & (1) Storage management issues, (2) Overhead analysis is not in details\\
\hline

Brotsis \textit{et al.} \cite{brotsis2019blockchain} & (1) Private evidence database is utilized along with a permissioned blockchain, (2) Provides security services like non-repudiation, integrity and authentication, (3) An intrusion detection tool is used, which helps in enhancing the security and determining the source of attacks, (4) Preserves and records the history of handling the evidence & (1) Lack of overhead and responsiveness analysis (2) Implementation details considering storage constraints \\
\hline

Novo \textit{et al.} \cite{novo2018blockchain} & (1) Offers various strengths such as transparency, auditability, decentralized consensus and security, (2) Provides scalable, generic and manageable access control system, (3) Due to the versatility of management hub nodes, high flexibility of the network is ensured & The network suffers from overhead of waiting due to the time required to issue control information by the blockchain network\\
\hline

Pourvahab \textit{et al.} \cite{pourvahab2019efficient} & (1) Lightweight forensic architecture to increase the security, integrity and assist CoC, (2) IoT devices are connected via gateway nodes and based on flow rules and action for each packet, (3) Increase in performance with minimal delay, overhead and processing time, (4) Increase in throughput, security and accuracy & (1) Lack of implementation details and storage overhead analysis\\
\hline

Xie \textit{et al.} \cite{xie2019blockchain} & (1) Accountability of the message is validated by using the immutability feature of blockchain, (2) Trust management is established so that messages cannot be tampered, (3) Ensures feasibility, accuracy and efficiency & (1) Overhead is introduced due to the encryption and transmission of messages and videos, (2) Lack of details of response time and latency\\
\hline

\end{tabular} 
\label{table:iotbc_sec2}
\end{table*}

\subsection{Security}

Blockchain based communication architecture, NGBN framework, is designed in~\cite{lei2019next} based on Named Data Networking (NDN) by integrating the IoT and wireless communication network. It provides a scalable, robust, and secure network environment for other applications. Two application scenarios illustrated under NGBN -- cyberspace scenario which focuses on the digital life and society, where the blockchain offers virtual coins so that the data can be transmitted efficiently and economically, which are useful for smart transportation-related application. 
Using blockchain as the underlying communication architecture, a security auditing system is constructed in~\cite{cha2018iso}, which can predict and detect any anomalies to guarantee confidentiality of data. The proposed system issues alerts for security incidents, collects evidences for auditing purposes, and probes event trajectory via data analysis in order to track the involved devices. It uses decentralized system to prevent tampering of data and help in encryption and decryption of data records. By using smart contract scheme on a single blockchain node, the effectiveness and efficiency of the security records are examined.
The IoT system presented in~\cite{niu2019blockchain} is based on blockchain to secure data and support efficient outsourcing services. To ensure confidentiality of data as well as to secure the keys against key-leakage attack, a blockchain-based key aggregation encryption system, known as BAI-KASE, is proposed, so that the encrypted data can be shared and searched securely. 
After performance analysis, the results indicate that the proposed model is efficient in computational costs as well as communication overheads.

A framework in mobile IoT is designed in~\cite{haseeb2019intrusion}, which is based on Wireless Sensor Network (WSN) for secure routing, to improve data reliability, network lifetime, and network security against various malicious attacks. It uses shortest routing chains to efficiently use energy with an optimum decision. Two subcomponents are proposed to generate and maintain non-overlapping clusters and to secure end-to-end routing paths based on the blockchain architecture which leads to the decrease in communication costs and overheads under large network and increase security.
Fabric-IoT, an access control system based on ABAC and Hyperledger Fabric Blockchain framework, is proposed in~\cite{liu2020fabric}, which has a device authority model ensuring the security of the device accessed. It helps in maintaining high throughput ensuring data consistency.
A blockchain based IoT system is proposed in~\cite{srivastava2019light} to provide advance security and privacy to the remote patient monitoring (RPM) system. The network provides reliable data communication and stores the data over cloud securely using advanced cryptographic techniques such as add–rotate–xor (ARX) encryption. Using digital signatures and ring signature, the privacy can be improved. To increase security, double encryption scheme is used, which protects our data from third party intruders. 

A blockchain-based IoT system is proposed in~\cite{agrawal2018continuous} using the distributed and immutable nature of blockchain which helps in building trust and making the system robust, secure, and immune to failure. Crypto-token which is pre-generated using prediction models, like Long Short-Term Memory (LSTM), is used to provide continuous security with seamless user authentication.
A Global IoT Device Discovery and Integration (GIDDI) solution in~\cite{dawod2019advancements} consists of a GIDDI Marketplace that provides functionality required for query, integration, payment, IoT device registration, security, and a GIDDI blockchain which manages metadata. 
The blockchain stores metadata related to IoT devices, which provides scalability, authentication and security.
Wireless battery management systems (WBMSs) of~\cite{faika2019blockchain} provide secure communication with external devices and data. It is used in cyber-physical environment to solve wiring-harness issues faced by battery management systems (BMSs), and utilizes cloud computing and wireless IoT networks efficiently. It incorporates smart contracts, peer-to-peer transaction, demand side management, supply chain, and IoT's privacy and security. Due to the application of the private blockchain system, less energy is required. It provides a key management and access control, thereby reducing latency significantly compared to other platforms. The latency of the experiment can be further reduced if instead of using cloud-based blockchain server, local blockchain server is used.

FastPay, a payment method for blockchain-based edge-IoT platforms, is proposed in~\cite{hao2018fastpay}. This method is both secure and fast, and is based on smart contracts mechanism. On front-end IoT devices, digital payments are made while in the back-end, blockchain's distributed nature ensures the validity of the system. It protects the system from double-spending attacks.
A blockchain-based B-IoT system is presented in~\cite{huang2019b}, which utilizes credit-based PoW mechanism. It decreases power consumption while increasing the complexity of computing for malicious nodes. 
It improves security as well as transaction efficiency. To protect the confidentiality of data, a data authority management system is designed in~\cite{lee2018towards}, which regulates the access to data and protects it without affecting the performance of the IoT system. It faces various challenges such as quality control of sensor data as well as storing large amounts of data.
Blockchain technology is used to revolutionize IoT system for Machine-to-Machine (M2M) communications~\cite{lee2018towards}. One M2M is a global project which achieves interoperability with respect to the communication system, which can be achieved using blockchain system. Its decentralized and peer-to-peer nature help secure the system from external attacks and it is based on blind voting. The proposed system in~\cite{lee2018towards} is applied in the IoT service layer; and it helps in preserving privacy and it uses smart contracts in various IoT related applications.  
NormaChain, as proposed in~\cite{liu2018normachain}, provides security against collusion and cryptoanalysis, and ensures privacy. 

A proxy re-encryption scheme is proposed in~\cite{manzoor2019blockchain} in combination with the blockchain to tackle trust issues and scalability of the system, and to ensure the confidentiality of data. The atomized system stores the encrypted IoT data in a distributed cloud and helps in securely sharing the data by using dynamic smart contracts without the involvement of a third party.
A multi-robot path planning is deployed in~\cite{mokhtar2019blockchain}. The plan combines Internet of Robotic Things (IoRT) with Hyperledger Fabric. IoRT acts as a bridge between the IoT domain and robotics domain. Hyperledger Fabric is an enterprise-grade and permissioned blockchain platform. 
Hyperledger Fabric provides an extensible and elastic architecture solution for control system by adapting power-efficient mechanism known as ordering service. It shows potential to enable scalability and security for IoT use cases. 
The results obtained reveals that while making transactions, a minimum latency is observed compared to other public blockchain platforms.
To improve the enactment of the IoT, and to ensure privacy and security, the IoT is integrated with blockchain, as discussed in~\cite{muzammal2018study}. 
A Decentralized IoT Collectability Data Marketplace (DCDM) market model as presented in~\cite{nguyen2019enabling}, where security is maintained, and to boost performance, a reward mechanism is used. 
Interledger is referred to as the class of operations, that is stretched across two or more distributed ledger technologies (DLTs), as discussed in~\cite{paavolainen2019interledger}. Group of researchers demonstrated a way in which a resource-constrained IoT device interacts with an ``IoT-friendly'' ledger by using interledger as a gateway service that mediates the payment and access controls between different parties using private and public permission-less ledgers such as blockchain and Ethereum. Using hashed timelock agreements (HTLAs), it is seen that payment can be made in an auditable and secure manner.

A decentralized application, DApp, is proposed in~\cite{papadodimas2018implementation}, which ensures security and flexibility to the users transacting on peer-to-peer level using cryptocurrencies. 
An analysis done by a group of researchers is shown in~\cite{pervez2019blockchain}, which ensures the security between various users in the network without involving third parties. 
An IoT system is proposed in~\cite{roman2018wip}, which collects, publishes, sends, and stores all the relevant data for smart sensors. 
It is free, instantaneous and trustworthy as reliable, and immutable information can be securely stored and can be accessed anytime by various stakeholders. 
Cloud-based IoT systems introduce constrained interaction in~\cite{samaniego2016hosting} with actuators and sensors as well as network latency. An IoT management construct, which is software based, is presented, and it enables multi-tenancy and load distribution on constrained edge-devices. It evaluates that, by reducing the number of virtual resources used by the host, the delay is increased, and it shows the use of permission-based blockchain. It is believed that the security and safety of the system can be increased by using blockchain for storing data.

Various blockchain platforms such as Ethereum and Hyperledger fabric are compared in~\cite{sanju2018energy} based on their consumption of energy for real workloads and by analyzing the performance trade-offs so that it could be securely implemented in IoT devices. It is found that Hyperledger fabric consumes less energy compared to Ethereum. 
The mechanism, proposed in~\cite{tzianos2019hermes}, is used for trading IoT sensors to provide security against malicious attacks and preserve privacy of the parties.
A blockchain-based IoT system is proposed in~\cite{urmila2019comparitive}, which is applied in healthcare sectors and helps in increasing the security and robustness by utilizing cryptographic algorithm. It helps in providing trustworthy and peer-to-peer platform to manage data efficiently without involving a third party. Blockchain provides a common platform so that different IoT devices can communicate in a distributed manner securely and protects against malicious attacks. 

The 5G technology promises reliability, scalability, real-time applications, ubiquitousness, and cost efficiency. It helps in improving IoT devices and makes it more efficient along with the help of blockchain. Communication constraints faced by the blockchain are resolved by the 5G communication in~\cite{dinesh2019conforming}, where the security and privacy issues are resolved by using blockchain. It faces some issues such as limited storage capacity of IoT systems, poor scalability, time consumption while mining, etc. It also increases the traffic overhead. 
Blockchain-assisted privacy-preserving authentication system (BPAS) is proposed in~\cite{choo2020blockchain}, which helps in transmitting messages securely without involving a centralized third-party. Similarly, data sourcing, crowdsourcing, edge computing, and industrial IoT (IIoT) architectures are achieved in overcoming some of the issues faced by the blockchain when it is implemented on IoT devices.

A blockchain technology addresses the problems of IoT security as discussed in~\cite{rashid2019security}. A self-clustering methodology is proposed, which helps in clustering the IoT network into $K$ number of unknown clusters by using Particle Swarm Optimization (PSO) and Genetic Algorithms, that help in optimizing the lifetime of the network. 
Hyperledger Fabric that helps in authorization, authentication and verification. The local blockchain helps in providing framework to improve safety and ability; and the global blockchain is implemented to explore upper layer communications. Due to the multi-layer approach, the network load, delay, and computational load are reduced. Due to the peer-to-peer nature of the model, communication efficiency is increased along with the integrity and security.
A survey is conducted in~\cite{alkurdi2018blockchain}, where the security aspects of blockchain-based IoT system is observed. 
A blockchain-based IoT solution is presented in~\cite{brotsis2019blockchain}, which helps in dealing with preservation and collection of digital forensic evidences. A private evidence database is utilized along with a type of permissioned blockchain, which helps to provide security services like non-repudiation, integrity, and authentication. 
To identify malicious attacks, an intrusion detection tool is used, which helps in enhancing the security and determining the source of attacks. Cyber-Trust Blockchain (CTB) is built on top of Hyperledger Fabric, and it dematerializes the chain-of-custody (CoC) process by preserving and recording the history of handling the evidences.

\begin{table}[!t]
\caption{Contributions of Existing Works Integrating IoT and Blockchain, and Concentrating on Security}
\centering
\begin{tabular}{|p{0.7cm}|p{7.1cm}|}
\hline
\textbf{Item} & \textbf{Contributions} \\
\hline
\cite{lei2019next} & Blockchain-based communication architecture is based on Named Data Networking by integrating IoT systems.\\
\hline
\cite{cha2018iso} & Security auditing system is constructed to predict and detect any anomalies to guarantee confidentiality of data.\\
\hline
\cite{niu2019blockchain} & Uses blockchain to secure data and support efficient outsourcing services.\\
\hline
\cite{haseeb2019intrusion} & Secure routing improves data reliability, network lifetime, and network security against various malicious attacks.\\
\hline
\cite{liu2020fabric} & Has a device authority model ensuring the security of the device accessed.\\
\hline
\cite{srivastava2019light} & Provides advance security and privacy to remote patient monitoring (RPM) systems.\\
\hline
\cite{agrawal2018continuous} & Helps in building trust and making the system robust, secure, and immune to failure.\\
\hline
\cite{dawod2019advancements} & Provides functionality required for query, integration, payment, IoT device registration, security, and managing metadata.\\
\hline
\cite{faika2019blockchain} & Incorporates smart contracts, peer-to-peer transaction, demand side management, supply chain, and IoT's privacy and security.\\
\hline
\cite{hao2018fastpay} & The approach is fast and is based on smart contracts mechanisms. On front-end IoT devices, digital payments are made while in the back-end, blockchain's distributed nature ensures the validity of the system.\\
\hline
\cite{huang2019b} & Utilizes the credit-based PoW mechanism and decreases the power consumption.\\
\hline
\cite{lee2018towards} & Regulates the access to data and protects it without affecting the performance of IoT systems\\
\hline
\cite{manzoor2019blockchain} & Stores the encrypted IoT data in a distributed cloud and helps in securely sharing the data by using dynamic smart contracts.\\
\hline
\cite{mokhtar2019blockchain} & Combines Internet of Robotic Things with Hyperledger Fabric, and acts as a bridge between IoT and robotics domains.\\
\hline
\cite{nguyen2019enabling} & Security is maintained, and to boost performance, a reward mechanism is used.\\
\hline
\cite{paavolainen2019interledger} & Using hashed timelock agreements, it is seen that the payment can be made in an auditable and secure manner.\\
\hline
\cite{pervez2019blockchain} & Ensures security between various users in the network without involving third parties.\\
\hline
\cite{samaniego2016hosting} & Proposes software-based IoT management construct to enable multi-tenancy and load distribution on constrained edge-devices.\\
\hline
\end{tabular} 
\label{table:t3}
\end{table}

\begin{table}[!t]
\caption{Contributions of Existing Works Integrating IoT and Blockchain, and Concentrating on Security (contd.)}
\centering
\begin{tabular}{|p{0.7cm}|p{7.1cm}|}
\hline
\textbf{Item} & \textbf{Contributions} \\
\hline
\cite{tzianos2019hermes} & Focuses on trading IoT sensors to provide security against malicious attack and preserve privacy of the parties.\\
\hline
\cite{urmila2019comparitive} & Helps in increasing the security and robustness in healthcare sectors by utilizing cryptographic algorithms.\\
\hline
\cite{dinesh2019conforming} & Communication constraints faced by the blockchain is resolved by 5G communications.\\
\hline
\cite{choo2020blockchain} & Helps in transmitting messages securely without involving centralized third-party.\\
\hline
\cite{rashid2019security} & A self-clustering methodology is used in clustering the IoT network into $K$ number of unknown clusters.\\
\hline
\cite{brotsis2019blockchain} & A private evidence database is utilized along with a type of permissioned blockchain, which helps provide security services like non-repudiation, integrity, and authentication.\\
\hline
\cite{novo2018blockchain} & Provides a scalable, generic and manageable access control system by implementing PoC.\\
\hline
\cite{pourvahab2019efficient} & Proposes SDN-based IoT network by using peer-to-peer decentralized nature of blockchain.\\
\hline
\cite{xie2019blockchain} & The vehicular IoT environment in 5G-VANETs enabling SDN architecture using the immutability feature of blockchain.\\
\hline
\cite{demirkan2020blockchain} & Different ways of impacts of blockchain in auditing are highlighted.\\
\hline
\cite{gupta2021towards} & Focuses on the development of a marketplace framework that can address design challenges in IoT models.\\
\hline
\cite{giannoutakis2020blockchain} & Provides a suitable smart contract scheme for guaranteeing the integrity of IoT devices.\\
\hline
\cite{sun2021blockchain} & Designs a mechanism to select multiple IoT devices to enable the real-time policy decisions in a distributed way.\\
\hline
\cite{agyekum2021proxy} & Proposes a proxy re-encryption mechanism for secure sharing of data in cloud environments.\\
\hline
\cite{sifah2021blockchain} & Overcomes the security issues of provenance data by leveraging blockchain.\\
\hline
\cite{ding2020incentive} & Incentive-based mechanism to purchase more power for computation from server devices.\\
\hline
\cite{oktian2020borderchain} & The security protocol is ensured by using verified IoT devices in the communication with IoT gateways.\\
\hline
\end{tabular} 
\label{table:t4}
\end{table}

To ensure scalability for constrained IoT devices, blockchain is used in~\cite{novo2018blockchain}, as it offers various strengths such as transparency, auditability, decentralized consensus, and security. The work provides a scalable, generic, and manageable access control system. It also implements PoC which enables various devices to connect to the same network. High flexibility of the network is ensured but the network suffers from overhead of waiting due to the time required to issue the control information by the blockchain network.  
On the Software-Defined Networking (SDN)-based IoT network, a lightweight forensic architecture is developed in~\cite{pourvahab2019efficient} by using the peer-to-peer decentralized nature of blockchain, in order to increase the security, integrity, and assist Chain of Custody (CoC) for evidence collection effectively. The blockchain-based controller uses Linear Homomorphic Signature (LHS) algorithm to validate users. The results indicate an increase in performance with a minimal delay, overhead, processing time, and response time. There is also an increase in the throughput, security, and accuracy.
A blockchain-based security framework is proposed in~\cite{xie2019blockchain} for the vehicular IoT environment in 5G-VANETs enabling SDN architecture. 
Accountability of the message is also validated by using the immutability feature of blockchain. Using blockchains, trust management is established such that the messages cannot be tampered. Though the model ensures feasibility, accuracy, and efficiency, there is an overhead introduced due to the encryption and transmission of messages and videos.

In~\cite{demirkan2020blockchain}, the uses of blockchain in financial security, financial accounting, and cyber security are discussed. The different ways of impacts of blockchain in auditing are also highlighted. Authors in~\cite{demirkan2020blockchain} find that blockchain needs to be effectively applied into different directions of cyber security and accounting. The authors in~\cite{gupta2021towards} focus on the development of a marketplace framework that can address design challenges in IoT models, and accordingly propose a three-tier framework. The mechanism proposed in~\cite{giannoutakis2020blockchain} provides a suitable smart contract scheme for guaranteeing the integrity of IoT devices, and immutable and dynamic management of malicious applications. In~\cite{sun2021blockchain}, a blockchain ledger is established to make more blockchain nodes from IoT devices. Here, the authors design a mechanism that selects multiple IoT devices to enable the real-time policy decisions in a distributed way.

In~\cite{agyekum2021proxy}, a proxy re-encryption mechanism is proposed for secure sharing of data in cloud environments, where an edge node behaves like a proxy server to handle extensive computations. The work~\cite{sifah2021blockchain} designs a generic cloud-based IoT framework to overcome the security issues of provenance data by leveraging blockchain. The work~\cite{singh2021blockchain} discusses blockchain-related security enhancement solutions by highlighting key points required to develop different blockchain systems. The authors in~\cite{mohanta2020addressing} review the security and privacy related issues in IoT systems, and present some security solutions using blockchain technology. For blockchain platforms, the incentive mechanism is investigated in~\cite{ding2020incentive} to purchase more power for computation from server devices in order to take part in the mining. For the access control, BorderChain~\cite{oktian2020borderchain} is based on blockchain in IoT endpoints, where the security protocol is ensured by using verified IoT devices in the communication with IoT gateways. BorderChain also applies access tokens which are used by IoT services and users for any query to IoT resources.
Tables~\ref{table:t3} and~\ref{table:t4} highlight existing works which consider the integration of the IoT and blockchain concentrating on security.

\subsection{Privacy}

A remote patient monitoring (RPM) system is presented in~\cite{srivastava2019light}. Using digital signatures and Ring Signatures, the privacy is achieved to protect data from third party intruders. 
WBMSs proposed in~\cite{faika2019blockchain} ensure secure communication with external devices and data. 
The credit-based PoW mechanism in~\cite{huang2019b} builds DAG structured blockchain which overcomes various problems such as power intensiveness, privacy and low-throughput, and makes the system more efficient.
A scalable and trustworthy mechanism is proposed in~\cite{kataoka2018trust} which enforces traffic management by integrating blockchain, pseudo IoT applications and SDN at the edge networks, for verification and authentication purposes. It reduces the risk of IoT deployment by interacting various IoT devices, stakeholders, and networks to manage the traffic efficiently and ensure security, reliability, and information tracking by the use of peer-to-peer nature of blockchain. Various IoT services and devices are listed, which are circulated among the stakeholders to prevent traffic and reduce attacks on edge networks. 
Various lightweight IoT devices face the problem of storing the blockchain due to low storage capacity. A storage compression consensus (SCC) algorithm is designed in~\cite{kim2019scc} which overcomes the problem such that the devices compress the blockchain using this algorithm to increase storage capacity. It is observed that, by using SCC, the storage capacity is reduced by $63\%$ compared to the existing algorithm like Byzantine fault tolerance (BFT) which uses private blockchain. However, the proposed scheme increases the maintenance of the system by acquiring free storage spaces.

\begin{table}[!t]
\caption{Contributions of Existing Works Integrating IoT and Blockchain, and Concentrating on Privacy}
\centering
\begin{tabular}{|p{0.7cm}|p{7.1cm}|}
\hline
\textbf{Item} & \textbf{Contributions} \\
\hline
\cite{srivastava2019light} & Using digital signatures and Ring Signatures, privacy is achieved to protect data from third party intruder.\\
\hline
\cite{faika2019blockchain} & Ensures secure communications with external devices.\\
\hline
\cite{huang2019b} & Proposes credit-based PoW mechanism with DAG structured blockchain.\\
\hline
\cite{kataoka2018trust} & Enforces traffic management by integrating blockchain, pseudo IoT applications and SDN at the edge networks.\\
\hline
\cite{kim2019scc} & Devices compress the blockchain to increase storage capacity.\\
\hline
\cite{lee2018towards} & M2M communications are presented to secure the system and maintain the privacy from external attacks.\\
\hline
\cite{paavolainen2019interledger} & Mediates the payment and access controls between different parties, where the payment can be made in an auditable and secure manner.\\
\hline
\cite{dinesh2019conforming} & Uses the 5G communication with blockchain ensuring privacy.\\
\hline
\cite{kim2018brics} & Ensures the privacy for resource-constrained IoT devices using Multi-Party Computation and blockchain.\\
\hline
\cite{kim2018brics} & Ensures the privacy for resource-constrained IoT devices using Multi-Party Computation and blockchain.\\
\hline
\end{tabular} 
\label{table:t5}
\end{table}

M2M communications presented in~\cite{lee2018towards} secure the system from external attacks. It helps in preserving the privacy.
Interledger discussed in~\cite{paavolainen2019interledger} acts as a gateway service that mediates the payment and access controls between different parties, where the payment can be made in an auditable and secure manner.
In~\cite{roy2018blockchain}, various challenges as well as current prospects of blockchain are surveyed.
Communication constraints faced in~\cite{dinesh2019conforming} are resolved by the 5G communication using blockchain ensuring the privacy. However, it faces some issues such as limited storage capacity, poor scalability, time consumption, and it increases overhead in traffic.
BPAS proposed in~\cite{choo2020blockchain} helps in transmitting messages securely. 
Two protocols, BRICS-BB and BRICS-SSP, are introduced in~\cite{kim2018brics} for resource-constrained IoT devices using Multi-Party Computation (MPC) and blockchain. 
These protocols are applied in variety of cases. For example, BRICS-BB is applied on software updates and BRICS-SSP can be used for data privacy management, supply chain, and identity management.
A survey of blockchain of things (BCoT) is conducted in~\cite{dai2019blockchain} which provides an overview of both the blockchain technology and Internet of Things as well as the challenges addressed by the blockchain technology in IoT systems. Industrial applications of BCoT and issues faced by using blockchain technology for fifth generation in IoT are also highlighted.
Table~\ref{table:t5} highlights existing works which consider the integration of the IoT and blockchain concentrating on maintaining the privacy.

\subsection{Authentication}

An IoT device residing on blockchain uses smart contracts to provide authentication, integrity, and non-repudiation. Due to the problem faced by IoT devices using centralized platform, a decentralized peer-to-peer based blockchain network is implemented in~\cite{choi2018blockchain}, which helps in preventing data-tampering by intruders.
The GIDDI solution proposed in~\cite{dawod2019advancements} is IoT-scaled and IoT-owned. In this solution, the blockchain stores metadata related to IoT devices, which provides scalability, authentication, and security.
An update framework is proposed in~\cite{dhakal2019private}, which prevents attacks and ensures authentication and repudiation to the messages sent between the firmware update service and IoT devices.
A blockchain-based architecture is proposed in~\cite{guin2018ensuring} for edge IoT devices so that they can be uniquely identified through registering physically unclonable functions (PUF) attributes at the time of manufacturing. 
The dual-use of blockchain enhances the usability and reliability of the system, and counteracts the problems of cloning and counterfeiting. A low-cost communication protocol is also introduced to securely authenticate the local devices.

A trust list is proposed in~\cite{kataoka2018trust}, which enforces traffic management for verification and authentication purposes. 
A blockchain-based IoT system is proposed in~\cite{le2019lightweight}, which is used to validate blockchain data helping IoT devices with random nodes to confirm block headers without only depending on the central server. But the large size of the blockchain causes problem for IoT devices with low data storage capacity, memory, and bandwidth. The proposed system works on the top of blockchain protocols supporting IoT devices with low-power. It consists of a gateway which transfers messages from blockchain to IoT devices. It uses a probabilistic approach which is based on Bloom filters. 
Due to the high cost of deployment, IoT-based solutions are restricted but it can be overcomed by horizontal integration of IoT devices and the use of blockchain system. In this direction, a ticket-based verification protocol is implemented in~\cite{pouraghily2019lightweight}, which does not require the participation of IoT devices in complex transactions, and therefore it can be used in low-end IoT devices. It reduces the networking and processing overhead of IoT devices in the system.

A blockchain framework, ``Sensor-Chain'', is designed for IoT of mobile devices in~\cite{shahid2019sensor}. The proposed framework is lightweight and scalable so that the IoT's limited storage capabilities as well as computational power can be addressed. The designed framework reduces resource consumption and is capable in retaining information about IoT systems of mobile devices. Authentication is concentrated in this work. 
BPAS is proposed in~\cite{choo2020blockchain}, which helps in transmitting messages securely without involving a centralized third-party. 
The user authentication scheme proposed in~\cite{almadhoun2018user} enables the integration of blockchain and fog nodes. The proposed scheme uses smart contracts to authenticate users such that IoT devices can be accessed without involving a third party in a decentralised network. Fog nodes provide scalability by carrying out computation related tasks to communicate with blockchain and authenticate devices. The security analysis is also conducted, which shows that the proposed solution achieves security goals of integrity, confidentiality, and availability. The proposed mechanism provides authentication, data encryption, and security against malicious attacks.

\begin{table}[!t]
\caption{Contributions of Existing Works Integrating IoT and Blockchain, and Concentrating on Authentication}
\centering
\begin{tabular}{|p{0.7cm}|p{7.1cm}|}
\hline
\textbf{Item} & \textbf{Contributions} \\
\hline
\cite{choi2018blockchain} & Proposes a decentralized peer-to-peer network to prevent data-tampering by intruders.\\
\hline
\cite{dawod2019advancements} & The blockchain stores metadata related to IoT devices, which provides scalability, authentication, and security.\\
\hline
\cite{dhakal2019private} & Ensures authentication and repudiation to the messages sent between the firmware update service and IoT devices.\\
\hline
\cite{guin2018ensuring} & Edge IoT devices can be uniquely identified through registering PUF attributes at the time of manufacturing.\\
\hline
\cite{kataoka2018trust} & Enforces traffic management for verification and authentication purposes.\\
\hline
\cite{le2019lightweight} & Validates blockchain data with random nodes to confirm block headers without only depending on the central server.\\
\hline
\cite{pouraghily2019lightweight} & A ticket-based verification protocol is implemented, which does not require the participation of IoT devices in complex transactions.\\
\hline
\cite{shahid2019sensor} & The proposed framework is lightweight and scalable, and suitable for devices having limited storage capabilities.\\
\hline
\cite{choo2020blockchain} & Helps in transmitting messages securely without involving centralized third-party.\\
\hline
\cite{almadhoun2018user} & Fog nodes provide scalability by carrying out computation related tasks to communicate with blockchain and authenticate devices.\\
\hline
\cite{guo2019blockchain} & Practical Byzantine fault tolerance consensus algorithm is implemented in an optimized manner to construct a consortium blockchain.\\
\hline
\cite{lau2018blockchain} & The proposed software system consists of ADCP handler, software firewall, WebSocket server, and blockchain database.\\
\hline
\cite{maitra2020proof} & Reduces the latency in block validation and energy consumption.\\
\hline
\end{tabular} 
\label{table:t6}
\end{table}

A trusted and distributed system is proposed in~\cite{guo2019blockchain}, which is based on edge computing and blockchain in order to improve the authentication and efficiency. 
The Practical Byzantine fault tolerance (PBFT) consensus algorithm is implemented in an optimized manner to construct a consortium blockchain which is used for storing authenticated data and logs. It achieves traceability of terminals and guarantees trusted authentication. Edge computing is applied to provide edge authentication and name resolution using smart contracts. Asymmetric cryptography is used to prevent the nodes from being attacked. Caching strategy is proposed which outperforms edge computing in terms of hit ratio and average delay.
Using blockchain technology, we can digitally identify the IoT devices, and then authenticate these devices before it join an IoT network. Authenticated Devices Configuration Protocol (ADCP) helps in achieving the authentication process as presented in~\cite{lau2018blockchain}. The proposed software system consists of ADCP handler, software firewall, WebSocket server, and blockchain database. WebSocket server helps in synchronizing blockchain database. 
ADCP responds to the requests for connection from other authenticated devices. The software firewall helps in encryption of data automatically.
A consensus algorithm is proposed in~\cite{maitra2020proof} that reduces the latency in block validation and energy consumption. The proposed mechanism is evaluated over resource-constrained IoT devices.
Table~\ref{table:t6} highlights existing works which consider the integration of the IoT and blockchain concentrating on maintaining the authentication.

\subsection{Tamper-Proof and Efficiency}

A security auditing system is constructed in~\cite{cha2018iso} which prevents tampering of data and helps in encryption and decryption of data records. 
Fabric-IoT proposed in~\cite{liu2020fabric} makes uses of blockchain's decentralization, traceability as well as tamper-proof property. The proposed system has a device authority model which ensures the security of the device. 
Due to the problem faced by IoT devices in using centralized platform, a decentralized peer-to-peer based blockchain network is implemented in~\cite{choi2018blockchain}, which helps in preventing data-tampering by intruders. 
To update firmware and software, instead of delta updates, a private blockchain mechanism is proposed in~\cite{dhakal2019private}, which enhances the update performance and helps in integrity verification, by utilizing a tamper-proof server. The blockchain server maintains a record of metadata, checksum, etc. 

\begin{table}[!t]
\caption{Contributions of Existing Works Integrating IoT and Blockchain, and Concentrating on Tamper-Proof and Efficiency}
\centering
\begin{tabular}{|p{0.7cm}|p{7.1cm}|}
\hline
\textbf{Item} & \textbf{Contributions} \\
\hline
\cite{cha2018iso} & Prevents tampering of data and helps in encryption and decryption of data records.\\
\hline
\cite{liu2020fabric} & Has a device authority model which ensures the security of accessed devices.\\
\hline
\cite{choi2018blockchain} & Helps in preventing data-tampering caused by intruders.\\
\hline
\cite{dhakal2019private} & Enhances the update performance and helps in integrity verification by utilizing a tamper-proof server.\\
\hline
\cite{pervez2019blockchain} & With the application of DAG and its multi-forked structure, high scalability, efficient provenance and optimized validation are observed.\\
\hline
\cite{damianou2019architecture} & Information circularity is fulfilled and improved in a smart city.\\
\hline
\cite{xie2019blockchain} & Using blockchain, a trust management is established such that the messages cannot be tampered.\\
\hline
\cite{zhou2018beekeeper} & Allows servers to process data without learning from homomorphic computations on the encrypted data.\\
\hline
\end{tabular} 
\label{table:t7}
\end{table}

A shared distributed ledger ensures secure and tamper-proof transactions between various users in the network without involving third parties. With the application of DAG and its multi-forked structure, the high scalability, efficient provenance, and optimized validation are observed in~\cite{pervez2019blockchain}. It affects Key Performance Indicators (KPIs) of the logistics domain. Together the IoT and blockchain help in real-time monitoring of resources, and ensure tamper-proof and efficiency.
To address the constrained nature of the IoT, both blockchain and edge computing are used. It decreases device requirements for memory, and increases the performance. 
The purpose of information circularity is fulfilled in a smart city as data can be extracted to reduce the waste production and resource consumption, and increase the services, as discussed in~\cite{damianou2019architecture}.
In~\cite{xie2019blockchain}, by the use of blockchain, a trust management is established such that the messages cannot be tampered. Although the model ensures feasibility, accuracy, and efficiency, there is an overhead introduced due to the encryption and transmission of messages and videos. 
A blockchain-based IoT system, known as Beekeeper, is proposed in~\cite{zhou2018beekeeper}, that allows servers to process data without learning from homomorphic computations on the encrypted data. 
The data can be collectively verified even by public. Since all the data are recorded in the decentralized blockchain, computational resources and huge memory are not required, and therefore the proposed mechanism ensures privacy, security, and the resistance to tamper.
Table~\ref{table:t7} highlights existing works which consider the integration of the IoT and blockchain concentrating on tamper-proof and efficiency.

\subsection{Resource Management}

The survey~\cite{khor2021public} presents the classification of resource-constrained IoT devices and associated wireless communication protocols. The authors in~\cite{yao2019resource} discuss the cloud computing integrated with the blockchain platform for offloading computations from IIoT networks. Moreover, the resource management issues are studied between cloud service providers and miners in blockchain. The work~\cite{xu2020blockchain} discusses the importance of the blockchain in 6G communications and resource management for multiple applications in the IoT. A distributed matching mechanism is proposed in~\cite{yang2020distributed} to maximize the utilization of the resource-restricted fog computing architecture, that guarantees different mining requirements of fog nodes.
Table~\ref{table:t8} highlights existing works which consider the integration of the IoT and blockchain concentrating on resource management.

\begin{table}[!t]
\caption{Contributions of Existing Works Integrating IoT and Blockchain, and Concentrating on Resource Management}
\centering
\begin{tabular}{|p{0.7cm}|p{7.1cm}|}
\hline
\textbf{Item} & \textbf{Contributions} \\
\hline
\cite{khor2021public} & Presents the classification of resource-constrained IoT devices and associated wireless communication protocols.\\
\hline
\cite{yao2019resource} & Discusses the cloud computing integrated with the blockchain platform for offloading computations from IIoT networks.\\
\hline
\cite{xu2020blockchain} & Highlights the importance of the blockchain in 6G communications and resource management in IoT.\\
\hline
\cite{yang2020distributed} & Maximizes the utilization of resource-restricted fog computing architecture.\\
\hline
\end{tabular} 
\label{table:t8}
\end{table}


\begin{table*}
\caption{Integration of IoT with Blockchain Concentrating on Privacy}
\centering
\begin{tabular}{|p{2.8cm}|p{8.5cm}|p{4.5cm}|}
\hline
\textbf{Item} & \textbf{Strength} & \textbf{Weakness} \\
\hline

Kataoka \textit{et al.} \cite{kataoka2018trust} & (1) Enforces traffic management with the help of SDN, (2) Reduces the risk of IoT deployment, (3) Ensures security, reliability and information tracking by the use of peer-to-peer nature of blockchain, (4) Reduces attacks on edge networks & Implementation constraints related to storage management are not highlighted\\
\hline

Kim \textit{et al.} \cite{kim2019scc} & Storage compression based consensus algorithm & Increases the maintenance complexity of the system by acquiring free storage spaces\\
\hline

Lee \textit{et al.} \cite{lee2018towards} & (1) Secures the system from external attacks, (2) Preserves privacy in various IoT related applications & (1) Protocol overhead, (2) No details of responsiveness analysis with respect to different existing related mechanisms\\
\hline

Paavolainen \textit{et al.} \cite{paavolainen2019interledger} & (1) Enhance privacy by acting as a gateway service that mediates the payment and access controls between different parties, (2) Payment can be made in an auditable and secure manner & (1) Implementation details, (2) Lack of analysis of trade-off between privacy preserving implementation constraints and responsiveness of the proposed mechanism\\
\hline

Dinesh \textit{et al.} \cite{dinesh2019conforming} & Communication constraints is resolved by 5G communication & (1) Limited storage capacity, (2) Poor scalability, (3) Time consumption, (4) Increase of traffic overhead\\
\hline

Choo \textit{et al.} \cite{choo2020blockchain} & (1) Identity verification maintaining trust, (2) Transmits messages securely without involving centralized third-party & Implementation related to security and computational costs\\
\hline

Kim \textit{et al.} \cite{kim2018brics} & (1) Authenticity, privacy and integrity of data transactions in a lightweight manner, (2) Fast and does not introduce overhead & (1) Latency of transaction output is not highlighted, (2) Implementation details in different applications\\
\hline

Dai \textit{et al.} \cite{dai2019blockchain} & (1) Provides an overview of both blockchain technology and IoT along with the challenges, (2) Industrial applications and issues faced by blockchain technology for fifth generation in IoT & Lack of details in implementation direction\\
\hline

\end{tabular} 
\label{table:iotbc_priv}
\end{table*}


\begin{table*}
\caption{Integration of IoT with Blockchain Concentrating on Authentication}
\centering
\begin{tabular}{|p{2.8cm}|p{8.5cm}|p{4.5cm}|}
\hline
\textbf{Item} & \textbf{Strength} & \textbf{Weakness} \\
\hline

Dawod \textit{et al.} \cite{dawod2019advancements} & Stores metadata related to IoT devices which provides
scalability, authentication and security & Design complexity and performance issues are not highlighted\\
\hline

Dhakal \textit{et al.} \cite{dhakal2019private} & Ensures authentication and repudiation to the messages sent between the firmware update service and IoT devices & Lack of details of communication between devices\\
\hline

Guin \textit{et al.} \cite{guin2018ensuring} & (1) The registered device can be verified anywhere without actually locating the manufacturer, (2) The local dataset helps in authenticating local edge IoT devices robustly, (3) Enhances the usability and reliability of the system, (4) Counteracts the problems of cloning and counterfeiting, (5) Low-cost communication protocol to securely authenticate the local devices & (1) Protocol implementation overhead, (2) Storage management issues\\
\hline

Kataoka \textit{et al.} \cite{kataoka2018trust} & Reduces the risk of IoT deployment by interacting with various
IoT devices and stakeholders & Trust list implementation details and related overhead are not highlighted\\
\hline

Le \textit{et al.} \cite{le2019lightweight} & (1) Works on top of blockchain protocols supporting IoT devices with low-power, (2) Space and time-efficient & (1) Constraints regarding data storage capacity are not in details, (2) Protocol overhead and implementation cost\\
\hline

Pouraghily \textit{et al.} \cite{pouraghily2019lightweight} & (1) Does not require the participation of IoT devices in complex transactions, (2) Reduces the networking and processing overhead of IoT devices & Storage management is not highlighted, (2) Response time and latency are not in details\\
\hline

Shahid \textit{et al.} \cite{shahid2019sensor} & (1) Lightweight and scalable, (2) Limited storage capabilities and computational power are addressed, (3) Reduces resource consumption, (4) Capable in retaining information about IoT systems of mobile devices, (5) Extends with an integration of smart contract to deal with data loss when number of nodes in a cell becomes relatively low & (1) Lack of details of implementation constraints, (2) Latency analysis is not in details\\
\hline

Almadhoun \textit{et al.} \cite{almadhoun2018user} & (1) Transmits messages securely without involving centralized third-party, (2) Fog nodes provide scalability by carrying out computation related tasks for communication, (3) Achieves security goals of integrity, confidentiality and availability, (4) Provides authentication, data encryption and security against malicious attacks & (1) Lack of details of storage management, (2) Overhead complexity, (3) Delay and responsiveness are not highlighted\\
\hline

Guo \textit{et al.} \cite{guo2019blockchain} & (1) Practical Byzantine fault tolerance consensus algorithm is implemented in an optimized manner, (2) Achieves traceability of terminals and guarantees trusted authentication, (3) Edge computing is applied to provide edge authentication, (4) Asymmetric cryptography is used to prevent the nodes from being attacked, (5) The proposed caching strategy outperforms edge computing in terms of hit ratio and average delay & (1) Protocol overhead needs a detailed analysis, (2) Details in comparative analysis with respect to response time and latency are required\\
\hline

Lau \textit{et al.} \cite{lau2018blockchain} & (1) WebSocket server helps in synchronizing blockchain database, (2) Storing all transactions helps in future computation of the proposed mechanism, (3) Software firewall helps in encryption of data automatically & (1) Details of storage handling are not highlighted, (2) Implementation costs and constraints are not analyzed\\ 
\hline

\end{tabular} 
\label{table:iotbc_auth}
\end{table*}


\begin{table*}
\caption{Integration of IoT with Blockchain Concentrating on Tamper-Proof and Efficiency}
\centering
\begin{tabular}{|p{2.8cm}|p{8.5cm}|p{4.5cm}|}
\hline
\textbf{Item} & \textbf{Strength} & \textbf{Weakness} \\
\hline

Liu \textit{et al.} \cite{liu2020fabric} & (1) Makes uses of blockchain's decentralization, traceability as well as tamper-proof property, (2) Ensures security of the device accessed, using three smart contracts, (3) Helps in maintaining high throughput ensuring data consistency & (1) Detailed analysis of efficiency and overhead are not highlighted, (2) Storage management constraints\\
\hline

Choi \textit{et al.} \cite{choi2018blockchain} & Helps in preventing data-tampering by intruders & A simple approach that needs a thorough analysis of implementation details, security, responsiveness, storage management etc.\\
\hline

Dhakal \textit{et al.} \cite{dhakal2019private} & (1) Enhances the update performance, (2) Helps in integrity verification and updates by utilizing a tamper-proof server, (3) Blockchain server maintains a record of metadata, checksum etc. & (1) Storage issues are highly related but these are not in details with an appropriate storage management approach, (2) Lack of implementation overhead analysis of the proposed mechanism\\
\hline

Pervez \textit{et al.} \cite{pervez2019blockchain} & (1) Ensures secure and tamper-proof transactions between various users in the network without involving third parties, (2) High scalability, efficient provenance and optimized validation are observed, (3) Helps in real-time monitoring of resources and ensures non-repudiation,
privacy, accountability, predictive analysis, transparency and stakeholder visibility & (1) Implementation details are not present, (2) Time complexity and storage constraints are not highlighted\\
\hline

Damianou \textit{et al.} \cite{damianou2019architecture} & (1) Decreases device requirements for memory and increases the performance, (2) The purpose of information circularity is fulfilled in a smart city as data can be extracted to reduce the waste production and resource consumption & (1) Response time analysis is not highlighted, (2) Overheads of storage management and device communication are not in details\\
\hline

Xie \textit{et al.} \cite{xie2019blockchain} & (1) Trust management is established so that messages cannot be tampered, (2) Ensures feasibility, accuracy and efficiency & Overhead is introduced due to the encryption and transmission of messages and video\\
\hline

Zhou \textit{et al.} \cite{zhou2018beekeeper} & (1) Processes data without learning from using homomorphic computations on the encrypted data, (2) High computational resources and huge memory are not required, (3) Ensures resistance to tamper, privacy and feasibility & (1) Lack of implementation details with performance issues (2) Time complexity analysis\\
\hline

\end{tabular} 
\label{table:iotbc_tamperproof}
\end{table*}


\begin{table*}
\caption{Integration of IoT with Blockchain Concentrating on Auditable, Trustworthy and Robustness}
\centering
\begin{tabular}{|p{2.8cm}|p{8.5cm}|p{4.5cm}|}
\hline
\textbf{Item} & \textbf{Strength} & \textbf{Weakness} \\
\hline

Novo \textit{et al.} \cite{novo2018blockchain} & Scalable, generic and manageable access control system along with the implementation of PoC & (1) Storage management issues, (2) Analysis of ease of application\\
\hline

Lei \textit{et al.} \cite{lei2019next} & (1) Provides a scalable, robust and secure network environment, (2) Smart city scenario which is useful for smart transportation-related application & (1) Implementation complexity, (2) Response time analysis along with the security perspective needs to be presented in details\\
\hline

Agrawal \textit{et al.} \cite{agrawal2018continuous} & Makes the system robust, secure and immune to failure & Lack of details of the proposed system along with implementation cost\\
\hline

Guin \textit{et al.} \cite{guin2018ensuring} & (1) Registered devices can be verified anywhere without actually locating the manufacturer, (2) Local dataset helps in authenticating local edge IoT devices robustly & (1) Details of access control management are not highlighted, (2) Lack of delay analysis with response time management\\
\hline

Wrona \textit{et al.} \cite{wrona2019use} & (1) Civil-military cooperation to store metadata collected from IoT devices and sensors using Hyperledger Fabric, (2) Ensures feasibility, trustworthiness and reliability of the collected data from IoT devices & (1) Architectural details and complexity analysis, (2) Communication overhead and responsiveness need to be highlighted\\
\hline

Bajpai \textit{et al.} \cite{bajpai2019blockchain} & (1) Identifies key points while integrating IoT and blockchain, (2) Three computing aspects: communication, storage and processing, (3) Overcomes single point of failure faced by various IoT systems, (4) Guarantees trustworthiness, data immutability and traceability, (5) Provides an audit trail to different stakeholders & (1) Lack of details of implementation and its constraints, (2) Details of latency analysis are not focused\\
\hline

Urmila \textit{et al.} \cite{urmila2019comparitive} & (1) Protects against malicious attacks, (2) Ensures feasibility, fault tolerance, scalability and operation time & (1) Lack of details of communication overhead, (2) Lack of details of the proposed distributed mechanism\\
\hline

Siris \textit{et al.} \cite{siris2019iot} & (1) Ensures the confidentiality and integrity of IoT data, (2) Hash-lock and time-lock mechanisms are used to cryptographically link various trusted IoT resources access, along with for authorisation grants and blockchain payments & (1) Lack of operation time analysis, (2) No details of storage management and implementation complexity along with cost management\\
\hline

\end{tabular} 
\label{table:iotbc_auditable}
\end{table*}

\subsection{Auditable, Trustworthy, and Robustness}

Using HTLAs in~\cite{paavolainen2019interledger}, it is observed that payment can be made in an auditable and secure manner.
A decentralised blockchain technology is used along with constrained IoT devices as it offers various strengths, such as transparency, auditability, etc., which make blockchain an ideal component of IoT solutions~\cite{novo2018blockchain}. In this work, a scalable, generic, and manageable access control system is proposed, which also implements PoC.
NGBN framework is designed in~\cite{lei2019next}, which provides a scalable, robust, and secure network environment for other applications, and it ensures trustworthy and robustness. It focuses on the digital life and society, where the blockchain offers virtual coins such that the data can be transmitted efficiently and economically. 
A blockchain-based IoT system is proposed in~\cite{agrawal2018continuous}, which helps in building trust and makes the system robust, secure, and immune to failure. 
In~\cite{guin2018ensuring}, by accessing the global dataset, the registered device can be verified anywhere without actually locating the manufacturer. The local dataset helps in authenticating local edge IoT devices robustly.

A blockchain-based IoT system is proposed in~\cite{wrona2019use}, which helps civil-military cooperation to store metadata collected from IoT devices and sensors, using Hyperledger Fabric. It ensures feasibility, trustworthiness, and reliability of the collected data from IoT devices. A high-level architecture is presented in~\cite{cha2018iso}, which helps in metadata binding using blockchain.
A decision matrix or an analysis framework is formulated in~\cite{bajpai2019blockchain}, which helps in identifying key points while integrating the IoT and blockchain. 
It guarantees trustworthiness, data immutability, and traceability and provides an audit trail to different stakeholders.
A blockchain-based IoT system helps in providing trustworthy and peer-to-peer platform to manage data efficiently without the involvement of a third party, as discussed in~\cite{urmila2019comparitive}. It provides a common platform such that different IoT devices can communicate in a distributed manner securely and protect against malicious attacks. It ensures feasibility, fault tolerance, scalability, and operation time. 

A model is proposed in~\cite{siris2019iot} to provide IoT devices the trust while making the payment by using blockchain and OAuth 2.0 authorization framework. It ensures the confidentiality and integrity of IoT data. Authorization requests are granted using the smart contract which works on permissioned blockchain. Hash-lock and time-lock mechanism are used to cryptographically link various trusted IoT resources for authorization grants and blockchain payments.
In~\cite{panarello2018blockchain}, a systematic survey is conducted on the integration of the IoT and blockchain, where the current trends are studied on the usage of blockchain in IoT systems. Challenges faced by IoT and blockchain are also highlighted and analyzed, directing to future works in the integration of the IoT and blockchain.

\section{Integration of IoT with AI}
\label{sec:IoT_AI}

In this section, we present the integration of the IoT and AI in details. We also discuss the limitations of the existing research works dealing with the applications of AI in the IoT. Table~\ref{table:iot_ai_security} and Table~\ref{table:iot_ai_privacy} highlight the strengths and weaknesses of the related works which consider the integration of the IoT with AI.

ML applied on inexpensive and low-power IoT devices is used for noise classification in~\cite{alsouda2018machine}, by using a supervised learning algorithm. Mel-frequency cepstral coefficients (MFCC) is used for audio feature extraction. To estimate the parameters for classification, Support Vector Machine (SVM) and K-Nearest Neighbours (KNNs) are used. The results show a high accuracy in the range $85\%-100\%$ with the high computational speed. It is observed that SVM has the highest accuracy. 
Using ML methods, IoT botnets can be detected; however, to minimize the number of features, the feature selection is applied in~\cite{bahcsi2018dimensionality}. It helps to overcome computation resource as well as scalability problem by increasing the accuracy rate based on the decision tree and a shallow method using the multi-class classifier. It can be readily interpreted by analysts and can be used in detecting the intrusion in the system.
A frost prediction system is proposed in~\cite{diedrichs2018prediction} enabling IoT devices. Using past readings of temperature and humidity, frost readings are predicted using ML algorithms. Data was augmented by Synthetic Minority Oversampling Technique (SMOTE). Experiments show a reduction in the prediction error and increase the performance in both Logistic Regression models and Random Forest by including the neighbour information. 

In~\cite{endler2017towards}, a semantic model is proposed for real-time processing and data stream processing based on Semantic Stream and Fact Stream. It assumes that all the objects, people, etc., have sensors embedded in them, that help to emit events whenever an action is performed. The main advantage is that it considers time as a key relation between different information. Using Complex Event Processing (CEP), the processing of streams can be implemented and it can be applied to any random Data Stream Management System (DSMS).
In~\cite{zantalis2019review}, a review is conducted to obtain the trend of IoT applications and ML techniques in Intelligent Transportation Systems (ITSs). It is seen that still there is a lack of application of ML techniques in smart parking applications and smart lighting systems. 
In~\cite{kanawaday2017machine}, the usage of Autoregressive Integrated Moving Average (ARIMA) forecasting in the IIoT is explored based on the time series data to predict the quality of defects and the possible failures. Hence, the mechanism improves the manufacturing process. The quality is managed and controlled by using ML, and the cost of maintenance is lowered. Different supervised models are used to evaluate the prediction accuracy and it is observed that they give an average accuracy of $0.96$.

A survey conducted in~\cite{mohammadi2018deep} reviews the characteristics of challenges faced by the IoT data for Deep Learning (DL) methods. The IoT big data as well as the IoT fast and streaming data are highlighted as the two categories of the IoT data generation and their analytic requirements. Several open source frameworks are presented for developing a DL architecture. Various challenges and the direction of future work are presented in~\cite{mohammadi2018deep} in the path of IoT applications using DL methods.
A monitoring system is developed in~\cite{patil2016early}, which helps in identifying grape diseases in the early stages using Hidden Markov Model. Based on the prediction done by using ML algorithms, alerts are sent to farmers and experts via SMS which helps them to increase the profits, protect the vineyards from diseases and reduce the manual efforts made for detecting the same. By following the irrigation schedule and correctly spraying the fertilizers, the excessive use of pesticides is limited. Hence, the proposed system increases the quality of the crop.
A survey is provided in~\cite{rodrigues2019machine} on the use of ML algorithms in Mobile Edge Computing (MEC) systems to learn about the problem, and find quick and optimal solutions for the same. It also lists the challenges while integrating them. Using ML, complex IoT systems can be built, which result in high quality services and supporting large number of resources. 
A survey is conducted in~\cite{shanthamallu2017brief} on the usage of ML algorithms in context of the IoT. Different types of ML methods are described in the field of sensor networks, health monitoring, IoT, pattern recognition, and anomaly detections. Various applications of ML algorithms in the field of the IoT-based healthcare are reviewed in~\cite{bharadwaj2021review}, where uses of various healthcare-based IoT applications are analyzed along with their possible improvements.

IoT devices are expected to support various resource-constrained Machine-Type Communication (MTC) devices by satisfying their QoS requirements and several other challenges faced by them, such as massive Machine Type Communications (mMTC) traffic characterization, small data packet transmission, QoS provisioning, and Radio Access Network (RAN) congestion. A review is included in~\cite{sharma2019toward} to address aforementioned issues. It identifies the potential as well as some challenges involved in using ML techniques to solve the issues.
Due to the large amount of data generated by IoT devices, large scale ML systems are required to organize, analyze, and draw inferences from the sensor IoT data. ML algorithms are data intensive and poses computational challenges to many computing devices including cloud. To bridge the efficiency gap, three approaches are highlighted in~\cite{venkataramani2016efficient}, which are approximate computing, machine learning accelerators, and post-Complementary Metal Oxide Semiconductor (CMOS) technologies. 
Various techniques are presented in~\cite{zhang2017machine} to implement Deep Neural Networks (DNNs) on Field Programmable Gate Arrays (FPGAs) having high energy efficiency and performance, which include the use of resource modelling, DNN reduction and re-training, configurable DNN, and resource allocation across DNN layers, in order to adapt the restricted resources used by IoT devices. A resource allocation strategy is presented, which drives theoretical guidelines for the minimal overall latency. Several designs are also showcased, such as Inception module for FaceNet face recognition, Long-term Recurrent Convolution Network (LRCN) for video captioning, and Long Short-Term Memory (LSTM) for sound recognition.

\begin{table}[!t]
\caption{Contributions of Existing Works Integrating IoT and AI}
\centering
\begin{tabular}{|p{0.7cm}|p{7.1cm}|}
\hline
\textbf{Item} & \textbf{Contributions} \\
\hline
\cite{alsouda2018machine} & Inexpensive and low-power IoT devices are used for noise classification using ML.\\
\hline
\cite{bahcsi2018dimensionality} & Overcomes computation resource as well as scalability problem by increasing the accuracy rate based on decision tree and a shallow method using multi-class classifier.\\
\hline
\cite{diedrichs2018prediction} & Frost prediction system is proposed using past readings of temperature and humidity.\\
\hline
\cite{endler2017towards} & A semantic model is proposed for real-time processing of data stream, based on Semantic Stream and Fact Stream.\\
\hline
\cite{zantalis2019review} & A review is conducted to obtain the trend of IoT applications and ML techniques in ITS.\\
\hline
\cite{kanawaday2017machine} & Improves the manufacturing process in IIoT by the usage of ARIMA forecasting in order to predict the quality of defects and the possible failures.\\
\hline
\cite{mohammadi2018deep} & Reviews the characteristics of challenges faced by IoT data for deep learning methods.\\
\hline
\cite{patil2016early} & Helps in identifying grape diseases in the early stages by using Hidden Markov Model.\\
\hline
\cite{rodrigues2019machine} & Learns about the problem and finds quick and optimal solutions for the use of ML algorithms in Mobile Edge Computing.\\
\hline
\cite{shanthamallu2017brief} & Different types of ML methods are described in the field of sensor networks, health monitoring, pattern recognition, and anomaly detection in IoT.\\
\hline
\cite{bharadwaj2021review} & Various healthcare-based IoT applications are analyzed along with their possible improvements.\\
\hline
\cite{sharma2019toward} & Reviews the issues related to resource-constrained MTC devices and mMTC traffic characterization, by satisfying their QoS requirements.\\
\hline
\cite{venkataramani2016efficient} & Highlights approaches to bridge the efficiency gap between IoT devices and ML algorithms.\\
\hline
\cite{zhang2017machine} & Implements DNNs on FPGAs having high energy efficiency and performance which includes the use of resource modelling.\\
\hline
\cite{cui2018survey} & Highlights the recent progress achieved by using ML techniques for IoT enabling users to develop intelligent IoT applications.\\
\hline
\cite{goap2018iot} & An open source architecture is proposed to predict the irrigation requirements by sensing the ground parameters such as soil temperature, soil moisture, environmental conditions, etc.\\
\hline
\cite{ventura2014ariima} & Reports usage patterns of Internet connected appliances by using ML algorithms, and processes these data in the cloud.\\
\hline
\end{tabular} 
\label{table:t9}
\end{table}

In~\cite{cui2018survey}, a comprehensive survey is conducted providing a view on the applications of ML in context of the IoT. It highlights the recent progress achieved by using ML techniques for the IoT enabling users to develop intelligent IoT applications and obtain deep analytics. It also discusses about the applications of the IoT using ML in different fields.
In~\cite{goap2018iot}, an open source architecture is proposed to predict the irrigation requirements by sensing the ground parameters such as soil temperature, soil moisture, environmental conditions, and weather forecast data retrieved from the Internet, by exploiting ML algorithms. 
In~\cite{ventura2014ariima}, a RESTful architecture is implemented on Internet connected appliances to report their usage patterns by using ML algorithms and the proposed mechanism processes these data in the cloud through Auto Regressive Integrated Moving Average (ARIIMA) predictive models. Energy consumption of these devices can be reduced in order to operate autonomously and efficiently. To address the issue of piglet mortality, the work~\cite{chen2020pigtalk} proposes PigTalk which is an AI-based IoT architecture for detecting and reducing piglet crushing, with the help of the real-time investigation of the collected voice data.
Table~\ref{table:t9} highlights existing works which generally consider the integration of the IoT and AI.


\begin{table*}
\caption{Integration of IoT with AI Concentrating on Security}
\centering
\begin{tabular}{|p{2.8cm}|p{8.5cm}|p{4.5cm}|}
\hline
\textbf{Item} & \textbf{Strength} & \textbf{Weakness} \\
\hline

Ventura \textit{et al.} \cite{ventura2014ariima} & (1) Demonstrates good transportability and is highly resilient with respect to external attacks, (2) $99\%$ of the devices were correctly identified from white list of IoT devices & Lack of convergence analysis in practical scenarios\\
\hline

Hussain \textit{et al.} \cite{hussain2020machine} & (1) A systematic review of the attack vectors, security requirements and the security solutions currently implemented for the constrained IoT networks, (2) Highlights various limitations such as fitting of models in all problems, convergence analysis, a minute change in the input creates havoc in the system and vulnerable to security breaches & Shortcomings of machine learning are not in details which are useful to determine the challenges of applying machine learning in IoT\\
\hline

Baracaldo \textit{et al.} \cite{baracaldo2018detecting} & (1) Discussion of the decrease of overall system performance due to the increase in poisoning attack, (2) Using tamper-free framework and contextual information, poisonous data is detected and filtered to train supervising learning model, (3) Increases performance and decreases runtime, (4) Ensures security & (1) Lack of analysis of learning overhead to ensure security, (2) Implementation details in practical perspective are not highlighted\\
\hline

Canedo \textit{et al.} \cite{canedo2016using} & (1) Anomalies in the data are detected by using artificial neural network, (2) Results correctly identify invalid data points in IoT systems & (1) Learning overhead is not in details, (2) Training constraints and future training perspective are not highlighted\\
\hline

Doshi \textit{et al.} \cite{doshi2018machine} & (1) DoS traffic attacks from customer IoT devices are identified
efficiently, (2) Computational overhead is restricted, (3) Limited feature set is used, which is necessary for middlebox deployment and real-time classification, (4) Five different classifiers are tested on both normal and DoS attack traffic dataset & (1) Learning convergence needs a detailed analysis, (2) Limitations with respect to security constraints are not highlighted\\
\hline

Li \textit{et al.} \cite{li2019system} & (1) Detects malicious behaviour and cyber-attacks in IoT devices
having limited communication and computation resources, (2) Can be implemented in any device despite the size, (3) More suitable for devices running specific type of applications & (1) More detailed implementation analysis is required to understand the dependency on the size and types of applications, (2) Tuning of different learning parameters is not highlighted\\
\hline

Xiao \textit{et al.} \cite{xiao2018iot} & (1) Learning based access control, secure offloading, IoT authentication and malware detection to protect data privacy, (2) Using transfer learning at the beginning of learning process reduces the risk of choosing bad security defence policies, (3) Low communication and computational overhead & A backup security solution should be incorporated and designed in order to provide reliability and security to the
system\\
\hline

Kumar \textit{et al.} \cite{kumar2018novel} & A three-tier architecture is proposed, which helps in storing and processing large volume of wearable sensor data & (1) Communication details and overhead between different tiers are not in details, (2) Lack of response time analysis and storage management\\
\hline

\end{tabular} 
\label{table:iot_ai_security}
\end{table*}


\begin{table*}
\caption{Integration of IoT with AI Concentrating on Privacy}
\centering
\begin{tabular}{|p{2.8cm}|p{8.5cm}|p{4.5cm}|}
\hline
\textbf{Item} & \textbf{Strength} & \textbf{Weakness} \\
\hline

Xiao \textit{et al.} \cite{xiao2018iot} & (1) Various attack models for IoT devices are investigated along with IoT security solutions based on ML techniques, (2) ML-based access control, secure offloading and IoT authentication, (3) Using transfer learning at the beginning of learning process reduces the risk of choosing bad security defense policies & (1) A backup security solution should be incorporated and designed in order to provide reliability and security, (2) Overhead of learning process, (3) Convergence details to impose security are not in details\\
\hline

Jeong \textit{et al.} \cite{jeong2017work} & (1) Helps in protecting the privacy of users whenever a resource-constrained IoT devices run ML applications, (2) To improve the privacy, partial processing is done at the client side so that it would be hard to retrieve the original data again, (3) Shorter prediction time compared to other local predictions & (1) There is a chance of increase in the time required for computation affecting the performance of the system, (2) Lack of learning overhead analysis\\
\hline

Li \textit{et al.} \cite{li2018learning} & (1) Due to the multiple layers in deep learning, accurate information can be extracted from the large amount of raw sensor data generated from IoT devices, (2) Edge computing
helps in reducing the traffic of the network, (3) Both deep learning and edge computing are utilized & (1) Since deep learning is used, learning time improvement mechanisms are required, (2) Storage utilization and implementation details are not highlighted\\
\hline

Pandey \textit{et al.} \cite{pandey2017machine} & Using a person's heart beat rate, it is tried to predict whether or not a person is in stress & (1) Details of improving the learning mechanism of the proposed model are not presented, (2) Lack of responsiveness analysis\\
\hline

\end{tabular} 
\label{table:iot_ai_privacy}
\end{table*}

\subsection{Security}

A supervised learning algorithm, Random Forest, is applied in~\cite{ventura2014ariima} to correctly identify types of IoT devices from the white list. To train the multi-class classifier, the data was manually labelled and collected from the traffic data. The results show that $99\%$ of the devices were correctly identified from white list. The proposed mechanism also demonstrates good transportability and it is highly resilient in respect to external attacks.
In~\cite{hussain2020machine}, a systematic review of the attack vectors, security requirements, and the security solutions currently implemented for the constrained IoT networks is presented. Here, ML and DL are used in the IoT devices to increase the security and privacy of the system. There are various limitations of using ML and DL, such as the model does not fit in all problems, the convergence takes long time, a minute change in the input creates havoc in the system, and the models are also vulnerable to security breaches. 
Due to the increase in the poisoning attack as described in~\cite{baracaldo2018detecting}, the overall performance of the system decreases and presents various security challenges. Using the tamper-free framework and contextual information, the poisonous data is detected and filtered to train the supervised learning model. Two variations for fully untrusted and partially trusted data sets are compared based on the performance. It is found that, by using the proposed method, the network performance is increased and the security is ensured.

\begin{table}[!t]
\caption{Contributions of Existing Works Integrating IoT and AI, and Concentrating on Security}
\centering
\begin{tabular}{|p{0.7cm}|p{7.1cm}|}
\hline
\textbf{Item} & \textbf{Contributions} \\
\hline
\cite{ventura2014ariima} & Identifies types of IoT devices from the white list using multi-class classifier.\\
\hline
\cite{hussain2020machine} & A systematic review of the attack vectors, security requirements, and security solutions for the constrained IoT networks is presented.\\
\hline
\cite{baracaldo2018detecting} & Using tamper-free framework and contextual information, poisonous data is detected and filtered to train supervising learning model.\\
\hline
\cite{canedo2016using} & By using artificial neural networks, anomalies in the data are intelligently detected.\\
\hline
\cite{doshi2018machine} & Performs DoS attack detection by using ML.\\
\hline
\cite{li2019system} & Detects malicious behaviour and cyber-attacks in IoT devices having limited communication and computation resources.\\
\hline
\cite{xiao2018iot} & Various attack models for IoT devices are investigated along with IoT security solutions based on ML techniques.\\
\hline
\cite{kumar2018novel} & Helps in storing and processing a large volume of wearable sensor data in IoT.\\
\hline
\cite{zhou2021secure} & Addresses the issues with the security and privacy in the edge intelligence model for IoT.\\
\hline
\cite{uprety2020reinforcement} & Presents reinforcement and deep learning based solutions to combat different types of security attacks in IoT models.\\
\hline
\cite{xu2020tt} & Designs a two-way trustworthy communication model in AI-enabled IoT frameworks.\\
\hline
\cite{makkar2020efficient} & The security of IoT devices is achieved by detecting spam with the help of ML, where five ML-based schemes are evaluated using different metrics.\\
\hline
\end{tabular} 
\label{table:t10}
\end{table}

To improve the security of IoT devices, a machine learning algorithm is used in~\cite{canedo2016using} within the IoT gateway. By using artificial neural networks, we can detect anomalies in the data intelligently. The results presented in~\cite{canedo2016using} correctly identify invalid data points in IoT systems.
By using ML, Denial of Service (DoS) detection is performed in~\cite{doshi2018machine}, where it is observed that DoS traffic attack can be identified efficiently from customer IoT devices. To restrict the computational overhead, limited feature set is used, which is necessary for the middlebox deployment and real-time classification. Five different classifiers, i.e., KNN, Support vector machine with linear kernel (LSVM), Decision tree using Gini impurity scores (DT), Random Forest using Gini impurity scores (RF), and Neural Network (NN) are tested on both normal and DoS attack traffic datasets. The results show an accuracy of higher than $0.99$ on all the aforementioned algorithms.  
An anomaly detection framework is proposed in~\cite{li2019system} for detecting malicious behaviours and cyber-attacks in IoT devices having limited communication and computation resources. The proposed mechanism can be implemented in any device despite the size and is more suitable for devices running specific type of applications. ML models and statistical techniques can be used in detecting any type of attacks on IoT devices, and the behavioural models of these IoT devices can be learned easily. It is seen that the proposed approach performs better in comparison to other traditional models. 

Various attack models for IoT devices are investigated in~\cite{xiao2018iot} along with IoT security solutions based on ML techniques which are learning-based including access control, secure offloading, IoT authentication, and malware detection to protect data privacy. The use of the transfer learning at the beginning of the learning process reduces the risk of choosing bad security defence policies. New ML techniques having low communication and computational overhead enhance the security of IoT systems. A backup security solution should be incorporated and designed in order to provide reliability and security to the system.
A three-tier architecture is proposed in~\cite{kumar2018novel}, which helps in storing and processing a large volume of wearable sensor data. Tier-1 helps in extracting data from IoT devices. Tier-2 stores a large volume of data in cloud. Tier-3 develops a logistic regression-based model for heart diseases.

In order to ensure the flexibility of data access and data security in the IoT, the scheme proposed in~\cite{zhou2021secure} addresses the issues with the security and privacy in the edge intelligence (EI) model for the IoT, where ML models are trained in EI. The work~\cite{uprety2020reinforcement} presents a survey of various classes of cyber-attacks in different IoT systems. Moreover, this work presents reinforcement and deep learning based solutions to combat different types of security attacks in several IoT models. The authors in~\cite{xu2020tt} design a two-way trustworthy communication model in AI-enabled IoT frameworks. The model incorporates both rating and trust information more thoroughly, which can alleviate the sparse trust problem in the IoT. In~\cite{makkar2020efficient}, the security of IoT devices is achieved by detecting spam with the help of ML, where five ML-based schemes are evaluated using different metrics considering a large set of input features. Each model calculates a spam score considering the input features, and this score reflects the trustworthiness of devices under various parameters.
Table~\ref{table:t10} highlights existing works which consider the integration of the IoT and AI concentrating on the security.

\subsection{Privacy}

Various attack models for IoT devices are presented in~\cite{xiao2018iot}, which address the IoT authentication and malware detection. The use of transfer learning at the beginning of learning process reduces the risk of choosing bad security defense policies. 
An approach is described in~\cite{jeong2017work}, which helps in protecting the privacy of users whenever a resource-constrained IoT device runs ML applications based on NN using cloud computing. To improve the privacy, partial processing is done at the client side such that it would be hard to retrieve the original data again. It is observed that the proposed approach has a shorter prediction time compared to other local predictions. However, there is a chance in increasing the time required for the computation, which affects the performance of the system.

\begin{table}[!t]
\caption{Contributions of Existing Works Integrating IoT and AI, and Concentrating on Privacy}
\centering
\begin{tabular}{|p{0.7cm}|p{7.1cm}|}
\hline
\textbf{Item} & \textbf{Contributions} \\
\hline
\cite{xiao2018iot} & Addresses IoT authentication and malware detection using transfer learning.\\
\hline
\cite{jeong2017work} & Helps in protecting the privacy of users whenever a resource-constrained IoT device runs ML applications using cloud computing.\\
\hline
\cite{li2018learning} & Using deep learning, the number of tasks in edge servers are risen along with achieving QoS requirements.\\
\hline
\cite{pandey2017machine} & Predicts whether or not a person is in stress, by exploiting both IoT and ML.\\
\hline
\end{tabular} 
\label{table:t11}
\end{table}

To maintain user privacy in the IoT platform, deep learning as well as edge computing are used in~\cite{li2018learning}. Due to the multiple layers in deep learning, the accurate information can be extracted from the large amount of raw sensor data generated from the IoT devices. Edge computing helps in reducing the traffic of the network to cloud servers. From the performance analysis, it is observed that the proposed solution can rise the number of tasks in edge servers along with achieving QoS requirements.
In~\cite{pandey2017machine}, using person's heart beat rate, the authors have tried to predict whether or not a person is in stress, by exploiting both the IoT and ML. The IoT informs patients about their stress conditions and ML algorithms are used to predict the condition of the patient. 
Table~\ref{table:t11} highlights existing works which consider the integration of the IoT and AI concentrating on the privacy.

\subsection{Link Quality, Fault detection, and Data Rate}

The work~\cite{dias2021prediction} proposes a methodology to dynamically estimate the link quality in cloud communications, considering locations of nodes. For this purpose, several machine learning techniques are implemented, such as decision tree, linear regression, neural networks, random forest, etc. Authors in~\cite{richardson2021towards} show that fault detection in data driven predictive analysis in remote locations can be achieved using the support vector machine algorithm and low bandwidth IoT. The scheme proposed in~\cite{mahesh2020data} utilizes the self-organization based map neural network to group IoT devices considering their data rates. For each of the group, a slot is allocated for channel access.

\subsection{Sentiment Analysis}

A learning-based mechanism is proposed in~\cite{yang2020fslm} to analyze sentiment of text in IoT frameworks, and for this purpose, a self-attention prototype is used to predict sentiment features. The work~\cite{dwivedi2018internet} discusses sentiment analysis of big data considering its features, definition, and decision value. Here, the challenges of the impact of the IoT on sentiment analysis of big data are also highlighted.

\section{Integration of IoT with Blockchain and AI}
\label{sec:IoT_AI_BC}

In this section, we discuss existing research works which integrate the IoT with blockchain and AI technologies. In this regard, the limitations of the related works are also discussed. Tables~\ref{table:iot_ai_bc} and~\ref{table:iot_ai_bc2} highlight the strengths and weaknesses of the related works which consider the integration of the IoT with blockchain and AI.

Sharing of authentic information using AI-capable machines in the IoT is very crucial for the risk management with personal accountability. A probabilistic proof system is introduced in~\cite{wohlgemuth2019competitive} using blockchain for cybersecurity satisfying economic and legal demands of the real-world business. The decentralized system helps in proving the trustworthiness of partners by a secure search. Here, the minimal external attacks or losses are expected due to the nonparticipation of third parties. The symmetric distribution of authentic information as well as overcoming the vulnerabilities of the adverse selection in the IoT contribute to an open marketplace on using security reports for trading rights~\cite{wohlgemuth2019competitive}.
The IoT is a transformational technology that minimizes human intervention with the aid of modern technologies like AI and blockchain. In spite of authentication of devices, the security remains a major issue which can be improved by using blockchain's decentralized and atomizing mechanisms, and transactions. Problems of network traffic, identity authentication, etc., can be overcomed by using smart contracts which also prevent tampering of data. Blockchain-based IoT systems can be used for supply chain management supporting access management systems, authorized payment generation, etc. AI, on the other hand, uses machine learning algorithms to adapt accordingly. Few concerns are that the blockchain system is not scalable and there is a lack of challenges for developing IoT applications~\cite{8649535}.

The MEC-based sharing system offers spatio-temporal services without the need of a central verification using smart contracts, as discussed in~\cite{rahman2019blockchain}. AI and cognitive computing are successful in providing reasoning by following the workflows and identifying or validating any changes found. Data transactions are advanced via blockchain, IoT, 5G device-to-device (D2D) communication capability, MEC, and fog computing. Phenomena of interest can be found by analyzing massive amount of data with the support of different types of data science advancements. Blockchain can store immutable data securely in a decentralized repository which can be shared using global identity, and has the potential to detect and eliminate fraudulent activities~\cite{rahman2019blockchain}. 
A software artifact, known as ``Smart Things''~\cite{samaniego2017internet}, is supported with AI features and is integrated with C Language Integrated Production System (CLIPS) to perform data analysis using RESTful micro services which make it self-inferencable and self-monitorable. ``Smart Things'' manages the real-time communication with other resources via Multichain, a permission-based blockchain protocol, to reduce the communication cost. A decentralized fog architecture is designed and implemented to distribute management tasks. In~\cite{samaniego2017internet}, the results illustrate that, to enable autonomous features, hybrid solutions should be adopted in IoT networks. 

In~\cite{ozyilmaz2018idmob}, a data marketplace is implemented on a blockchain-based platform using Swarm as a distributed storage system, where technologies such as the IoT, blockchain, and AI can interact and collaborate to create a decentralized platform, such that data can be stored and accessed easily by multiple parties. Thus, the proposed mechanism increases the service quality. It uses a voting mechanism to eliminate unreliable data providers. Blockchain provides various economic, technical, and user-facing benefits but it does not support safety-critical applications due to the lack of some form of encryption. Data replication is to be handled by the vendors.
A paid-sharing architecture of the knowledge market is proposed in~\cite{lin2019making} for making the knowledge tradable using edge-AI enabled IoT devices, where blockchain helps in ensuring security as well as efficiency in the market using a new consensus mechanism, known as Proof-of-Trading (PoT), that consumes less resources. A game-based knowledge strategy is proposed as incentives for the market, where it is observed that an optimal pricing strategy helps buyers with the high quality knowledge and encourages sellers to learn more data. Using Machine Learning, a hierarchical market model is introduced, as the extracted knowledge can be reproduced by it with almost no costs~\cite{lin2019making}. 

Blockchain is used to facilitate illegal market activities. To overcome it, a self-redactable blockchain (SRB) is proposed in~\cite{huang2019achieving} for building a trust layer for the IoT, using revocable chameleon hash (RCH) to create disincentives to manage any data-driven business and support execution of chain redaction securely as well as efficiently. To obtain redaction, additional metadata and costs are introduced, which can be compensated by the merits of the fast verification and storage saving provided by the same~\cite{huang2019achieving}.
Smart logistics solution is proposed in~\cite{arumugam2018iot}, where logistics planner and smart contracts are used to secure trust, provide visibility, and help in condition-monitoring of assets using distributed ML in supply chain management. It demonstrates liability, transparency, accountability, optimal planning, and traceability, for handling assets by different parties involved in the logistics of supply chain management with the help of the IoT, blockchain, ML, and Big Data.

The IoT and CPS maintain visibility and global trackability of supply chain by gathering information in real time. Manufacturers of these smart devices apply their own standards making interoperability more challenging leading to an overcomplicated system. Various cybersecurity risks are present due to the increase in interconnectivity leading to security breaches in sectors like finance. A framework is proposed in~\cite{ahmadi2019federated} using the IoT and CPS objects to incorporate forensic-readiness for investigations and post-incident; and a prototype is maintained to detect anomaly with the help of Bluetooth proximity monitoring. The proposed scheme ensures confidentiality, integrity, availability, authenticity, and safety~\cite{ahmadi2019federated}. 
Using the IoT and AI, a proposition is made in~\cite{chandra2019blockchain} to operate Halal food chain on blockchain with respect to its architecture, application, and technological components to safeguard data. 
Blockchain reduces chances of tampering with data, facilitates audit, provides security and compliance but its immutable nature makes compliance to new data privacy challenging~\cite{chandra2019blockchain}.

\begin{table*}
\caption{Integration of IoT with Blockchain and AI}
\centering
\begin{tabular}{|p{2.8cm}|p{8.5cm}|p{4.5cm}|}
\hline
\textbf{Item} & \textbf{Strength} & \textbf{Weakness} \\
\hline

Wohlgemuth \textit{et al.} \cite{wohlgemuth2019competitive} & (1) Symmetric distribution of authentic information as well as overcoming the vulnerabilities of adverse selection in IoT, (2) Minimal external attacks or losses are expected due to the nonparticipation of third parties & Responsiveness of the blockchain architecture in the presence of learning overhead is not in details\\
\hline

Jindal \textit{et al.} \cite{8649535} & (1) Blockchain-based IoT systems support access management systems and identity automatic currency exchange systems, authorized payment generation etc., (2) AI strengthens the proposed mechanism to adapt accordingly by using machine learning algorithms & (1) Blockchain system is not scalable, (2) Lack of professionals for developing IoT applications\\
\hline

Rahman \textit{et al.} \cite{rahman2019blockchain} & (1) MEC-based sharing system offers spatio-temporal services
without the need of central verification using smart contracts, (2) AI and cognitive computing are useful in providing reasoning by following the workflows and identifying or validating any changes found & (1) Protocol overhead and learning convergence are not highlighted, (2) Lack of implementation details\\
\hline

Samaniego \textit{et al.} \cite{samaniego2017internet} & (1) Software artifact is supported with AI features and is integrated with CLIPS system to perform data analysis automatically, (2) Permission-based blockchain protocol to reduce the communication cost, (3) Decentralized fog architecture to distribute management tasks, (4) Manages real-time communication with other resources & (1) Training constraints and the scope of improvement of the learning are not in details, (2) Lack of details of the proposed approach\\
\hline

{\"O}zyilmaz \textit{et al.} \cite{ozyilmaz2018idmob} & (1) Voting mechanism to eliminate unreliable data providers, (2) Provide various economic, technical and user-facing benefits, (3) Blockchain and AI interact and collaborate to
create a decentralized and trustless platform & (1) Does not support safety-critical applications due to the lack of some form of encryption, (2) Data replication is to be handled by the vendors\\
\hline

Lin \textit{et al.} \cite{lin2019making} & (1) Blockchain helps in ensuring security as well as efficiency in the market, (2) Architecture of knowledge market is proposed for making knowledge tradable using edge-AI, (3) Consumes less resources, (4) Optimal pricing strategy helps buyers with high quality knowledge and encourages sellers to learn more data, (5) Knowledge can be reproduced with almost no costs & (1) Design complexity can increase the overhead of protocol implementation and execution, (2) Scope of improvement of the trained model is not highlighted\\
\hline

Huang \textit{et al.} \cite{huang2019achieving} & (1) Self-redactable blockchain is proposed for building a trust layer for IoT, (2) Execution of chain redaction securely as well as efficiently, (3) Fast verification and storage saving & (1) Additional metadata and costs are introduced, (2) Convergence of learning in practical scenarios is not in details, (3) Lack of responsiveness analysis\\
\hline

Arumugam \textit{et al.} \cite{arumugam2018iot} & (1) Helps in condition-monitoring of assets using distributed ML framework in supply chain management, (2) Demonstrates liability, transparency, accountability, optimal planning and traceability, for handling assets by different parties & (1) Storage management is not presented, (2) Lack of details of the execution flow of the proposed model, (3) Adaptability details and constraints are not highlighted\\
\hline

Ahmadi-Assalemi \textit{et al.} \cite{ahmadi2019federated} & (1) Incorporates forensic-readiness for investigations and post-incident, (2) A prototype is maintained to detect anomaly with the help of Bluetooth proximity monitoring, (3) Ensures confidentiality, integrity, availability, authenticity as well as safety & (1) Handling cybersecurity needs to be detailed, (2) Lack of implementation details of the prototype along with management of learning overhead\\
\hline

\end{tabular} 
\label{table:iot_ai_bc}
\end{table*}

\begin{table*}
\caption{Integration of IoT with Blockchain and AI (contd.)}
\centering
\begin{tabular}{|p{2.8cm}|p{8.5cm}|p{4.5cm}|}
\hline
\textbf{Item} & \textbf{Strength} & \textbf{Weakness} \\
\hline

Chandra \textit{et al.} \cite{chandra2019blockchain} & (1) An increase in trust throughout the supply chain is noted, (2) Blockchain reduces chances of tampering with data, facilitate audit, provide security and compliance &
(1) Immutable nature of blockchain makes compliance to new data privacy challenging, (2) Storage management and implementation constraints are not highlighted\\
\hline

Shen \textit{et al.} \cite{shen2019privacy} & (1) Using blockchain, a secure and reliable platform is built for sharing data between data providers, (2) Stores data in distributed ledger without involving a third-party, (3) Secure and efficient training algorithm, (4) Ensures confidentiality of data for data analysis & (1) Lack of details of communication overhead in data sharing, (2) Implementation constraints, (3) Convergence analysis of the learning, (4) Storage management issues are not highlighted\\
\hline

Wang \textit{et al.} \cite{wang2019video} & (1) Helps in acquiring large amount of information and processing redundant and private sensor data, (2) Blockchain guarantees reliability, security as well as the robustness of the system, (3) CNN is used for real-time monitoring of data, (4) Ensures the security of the data and tolerance, (5) IPFS and edge computing reduces the storage and transmission cost & (1) Detailed analysis of latency and responsiveness needs to be presented, (2) Protocol implementation complexity is not in details\\
\hline

Zhao \textit{et al.} \cite{zhao2019mobile} & (1) Based on reputation crowdsourcing and provides better services, (2) An incentive mechanism is designed to increase the participation of customers in crowdsourcing tasks, (3) ML helps in predicting the consumption behaviours and customers requirements of the system, (4) Blockchain helps in increasing the security and privacy of the system & (1) Communication overhead and delay in crowdsourcing are not highlighted, (2) Scopes of improvement in training and convergence are not in details\\
\hline

Chen \textit{et al.} \cite{chen2018iot} & (1) Helps to ensure privacy as swell as data fusing, (2) All the resources can be exchanged between clouds and blockchain through trusted environment, and transactions are recorded, (3) The use of smart contracts helps in providing trustworthy and impartial environment & (1) Lack of details of communication overhead in exchanging information, (2) Storage management issues are not highlighted\\
\hline

Ramachandran \textit{et al.} \cite{ramachandran2018development} & (1) Transparency and security features are achieved while transferring data, (2) Helps cars to drive automatically, (3) Installation cost is comparatively cheaper & (1) Learning convergence is not in details, (2) Scope of minimization in response time and latency\\
\hline

Singla \textit{et al.} \cite{singla2018machine} & (1) Smart home IoT system is used to generate customizations based on the activity prediction of users, (2) Distributed association rule mining is applied to generate the rules of the activities from the users' device logs & Details of the mechanism and implementation constraints are not highlighted\\
\hline

Outchakoucht \textit{et al.} \cite{outchakoucht2017dynamic} & (1) Ensures distributed aspect which in turn increases the privacy of the model, (2) Machine Learning focuses on the dynamic as well as optimized aspect of the model, (3) Helps in security & (1) Lack of details of the adaptability issues, (2) Responsiveness analysis is not in details \\
\hline

\end{tabular} 
\label{table:iot_ai_bc2}
\end{table*}

Based on blockchain, a privacy preserving SVM scheme, known as SecureSVM, is proposed in~\cite{shen2019privacy} to encrypt the IoT data. Using blockchain, a secure and reliable platform is built for sharing data between data providers, and data is stored in a distributed ledger without involving a third-party. By employing a homomorphic cryptosystem, a secure, efficient, and accurate training algorithm is constructed such that a secure comparison and polynomial multiplication are achieved, which help in ensuring the confidentiality of data for data analysis~\cite{shen2019privacy}. For smart homes, the privacy-preserving distributed mechanism proposed in~\cite{yang2021privacy} enables users to manage energy usages parallelly in an optimized way. Moreover, a blockchain system is designed for developing a smart contract for IoT devices to support an energy management system.

In~\cite{wang2019video}, a video surveillance system is proposed based on Inter Planetary File System (IPFS) technology, permissioned blockchain, convolution neural networks (CNNs), and edge computing. The edge computing helps in acquiring a large amount of information and processing redundant data. IPFS is used for storing massive video data. Blockchain guarantees the reliability, security as well as the robustness of the system, whereas CNN is used for real-time monitoring of data. The system ensures the security of the data and tolerance. IPFS and edge computing combinely reduce the storage and transmission costs~\cite{wang2019video}.

\begin{table}[!t]
\caption{Contributions of Existing Works Integrating IoT, Blockchain, and AI}
\centering
\begin{tabular}{|p{0.7cm}|p{7.1cm}|}
\hline
\textbf{Item} & \textbf{Contributions} \\
\hline
\cite{wohlgemuth2019competitive} & A probabilistic proof system is introduced using blockchain for cybersecurity satisfying economic and legal demands of real-world business.\\
\hline
\cite{rahman2019blockchain} & Data transactions are advanced via blockchain, IoT, 5G D2D communication capability, MEC, and fog computing.\\
\hline
\cite{samaniego2017internet} & Proposes a decentralized fog architecture that manages real-time communications and reduces the communication cost.\\
\hline
\cite{ozyilmaz2018idmob} & A data marketplace is implemented on a blockchain-based platform, where data can be stored and accessed easily by multiple parties.\\
\hline
\cite{lin2019making} & A paid-sharing architecture of knowledge market is proposed using edge-AI enabled IoT devices, where Proof-of-Trading (PoT) consumes less resources.\\
\hline
\cite{huang2019achieving} & A self-redactable blockchain is proposed for building a trust layer for IoT to manage any data-driven business and support execution of chain redaction securely as well as efficiently.\\
\hline
\cite{arumugam2018iot} & Smart logistics solution is proposed, where smart contracts are used to secure trust, provide visibility, and help in condition-monitoring of assets using distributed ML.\\
\hline
\cite{ahmadi2019federated} & A framework is proposed using IoT and CPS to incorporate forensic-readiness for investigations and post-incident, and a prototype is maintained to detect anomaly with the help of Bluetooth proximity monitoring.\\
\hline
\cite{chandra2019blockchain} & Using IoT and AI, a proposition is made to operate Halal food chain on blockchain with respect to its architecture, application, and technological components to safeguard data.\\
\hline
\cite{shen2019privacy} & By employing a homomorphic cryptosystem, a secure, efficient, and accurate training algorithm is constructed.\\
\hline
\cite{yang2021privacy} & A blockchain system is designed for developing a smart contract for IoT devices to support an energy management system.\\
\hline
\cite{wang2019video} & A video surveillance system is proposed based on Inter Planetary File System technology, permissioned blockchain, convolution neural networks, and edge computing.\\
\hline
\cite{zhao2019mobile} & ML helps in predicting the consumption behaviours and customers' requirements of the system, and blockchain helps in increasing the security and privacy of the system.\\
\hline

\end{tabular} 
\label{table:t12}
\end{table}

\begin{table}[!t]
\caption{Contributions of Existing Works Integrating IoT, Blockchain, and AI (contd.)}
\centering
\begin{tabular}{|p{0.7cm}|p{7.1cm}|}
\hline
\textbf{Item} & \textbf{Contributions} \\
\hline
\cite{zhao2020privacy} & Develops a FL system to assist home device manufacturers to prepare (train) a ML model based on data provided by customers.\\
\hline
\cite{zhang2020blockchain} & Presents a framework of blockchain-based FL systems for failure estimation in IIoT to ensure verifiable integrity of data.\\
\hline
\cite{chen2018iot} & Helps to ensure the privacy as swell as data fusing.\\
\hline
\cite{ramachandran2018development} & A system with an autonomous feature is proposed using Raspberry Pi, machine learning, and image processing.\\
\hline
\cite{singla2018machine} & Generates customizations based on the activity prediction of users for IoT devices.\\
\hline
\cite{outchakoucht2017dynamic} & Ensures distributed, dynamic, and optimized aspects of the model and increases the security of the system.\\
\hline
\cite{xu2020edgence} & Intelligently handles massive decentralized applications in IoT frameworks.\\
\hline
\cite{rahman2021smartblock} & An AI-based layered hierarchical model is proposed to deploy a distributed blockchain-enabled SDN framework for IoT.\\
\hline
\cite{hamdaoui2020iotshare} & Applies a self-recovery mechanism to impose robustness against maliciousness and device failures.\\
\hline

\end{tabular} 
\label{table:t13}
\end{table}

A federated learning (FL) system is presented in~\cite{zhao2019mobile} for IoT manufacturers, where the proposed system is based on reputation crowdsourcing and provides better services than the existing related functionalities. Various technologies, such as blockchain, distributed storage, mobile edge computing, and federated learning, are used to construct the system. An incentive mechanism is designed to increase the participation of customers in crowdsourcing tasks. ML helps in predicting the consumption behaviours and customers' requirements of the system with the help of the federated technology, whereas blockchain helps in increasing the security and privacy of the system.
To design a smart home model, authors in~\cite{zhao2020privacy} develop a FL system to assist home device manufacturers to prepare (train) a ML model based on data provided by customers. Here, Manufacturers can estimate customers' future consumption and requirement behaviors. The work~\cite{zhang2020blockchain} presents a framework of blockchain-based FL systems for failure estimation in IIoT, which ensures verifiable integrity of data. Moreover, to address the issue of data heterogeneity in IIoT, a federated averaging-based FL algorithm is proposed, that considers the distance between negative and positive classes of each data set.

An IoT service platform is proposed in~\cite{chen2018iot} based on Joint Cloud Computing (JCC) to help ensuring the privacy as swell as data fusing. All the resources can be exchanged between clouds and blockchain through a trusted environment, where all the transactions are recorded. All trades are done using smart contracts which help in providing trustworthy and impartial environments.  
To an existing electric car, a system with an autonomous feature is proposed in~\cite{ramachandran2018development} using Raspberry Pi, machine learning, and image processing. With the help of blockchain, transparency and security features are achieved while transferring data. The system helps the car to drive automatically and the installation cost of the proposed model is comparatively cheaper. 

In~\cite{mohanta2020survey}, a survey of different technologies, such as blockchain, ML, and artificial intelligence, is conducted in order to address various security issues such as layer-wise issues and Confidentiality, Integrity, Availability (CIA) faced in IoT systems. Several research challenges of these technologies are also highlighted in~\cite{mohanta2020survey}. 
By using Machine leaning models, a smart home IoT system is used in~\cite{singla2018machine} to generate customizations based on the activity prediction of users for IoT devices. In~\cite{singla2018machine}, blockchain is used such that the model can run in a decentralized and secure manner. The distributed association rule mining is also applied in order to generate the rules of the activities from the user's device logs. 
A fully distributed and dynamic security policy is proposed in~\cite{outchakoucht2017dynamic} focusing on the access control in the IoT. Blockchain ensures the distributed aspect which in turn increases the privacy of the model. Whereas, ML, particularly reinforcement learning, focuses on the dynamic and optimized aspect of the model, which helps in the security part of the system. In addition, the architecture of the proposed framework is well exposed and explained. 

To intelligently handle massive decentralized applications in IoT frameworks, Edgence~\cite{xu2020edgence} is designed to use edge and cloud computing to communicate with IoT devices and end-users. In this case, an in-built blockchain is used to realize self-supervision and self-governing of edge clouds. In~\cite{rahman2021smartblock}, an AI-based layered hierarchical model is proposed to deploy a distributed blockchain-enabled SDN framework for the IoT, that ensures secure network communication and cluster-head selection via the identification of rouge switches. The work~\cite{he2020blockchain} addresses the privacy and security challenges of edge-computing-based IoT, and accordingly highlights the characteristics of blockchain, which are suited for IoT scenarios. The proposed framework is based on deep reinforcement learning and specifies the procedure for transactions between an edge node and IoT device. In~\cite{zhang2020data}, to propose an approach for an effective identification of faults in the network, an IoT model and edge-data verification mechanism are designed using blockchain and random forest algorithm. The authors in~\cite{hamdaoui2020iotshare} design a protocol that relies on peer-to-peer communications to allow interactions among IoT devices organized in a distributed way. The protocol applies a self-recovery mechanism to impose robustness against maliciousness and device failures. The protocol also uses a reputation model to monitor the devices and the quality of their service delivery, such that service delivery reputations can be leveraged for the selection of future devices.

Tables~\ref{table:t12} and~\ref{table:t13} highlight existing works which consider the integration of the IoT with blockchain and AI.

\section{Challenges and Future Reseacrh Scopes in the Integration of IoT with Blockchain and AI}
\label{sec:challenge}

There are several issues in integrating blockchain and AI with the IoT, and these issues need to be addressed in order to fully exploit the utilities of blockchain and AI in the IoT paradigm. In this section, we present the challenges in applying blockchain and AI in the IoT to design secure, scalable, intelligent, and robust IoT systems. These challenges will open new research scopes in designing smart IoT models.



\subsection{Challenges in Application of Blockchain in IoT}

Different challenges of applying blockchain in IoT are discussed as given in the following.

\subsubsection{Storage-Constraint in Blockchain Implementation}

The storage issue becomes serious when blockchain is integrated with the IoT~\cite{reyna2018blockchain,samaniego2016hosting,zhou2018beekeeper}. With a high capacity of storage, blockchain nodes cannot handle the challenge of increasing data amount. Furthermore, the IoT platform can have heterogeneous devices having various storage capacities, and thus the storage issues become more complicated while blockchain is associated with the IoT~\cite{wu2019comprehensive}.
Therefore, the integration of blockchain and IoT devices with a low capacity will become a great challenge.
In the IIoT, the application of blockchain requires a massive storage space, which is a great challenge in a resource-constrained IIoT architecture. The work~\cite{wang2019chainsplitter} addresses this issue by proposing a hierarchical blockchain storage architecture where the major portion of the blockchain is stocked in the cloud. Since an edge-based architecture can provide a faster response time than a cloud-based structure, it will be beneficial to use the storage of an edge architecture to implement blockchain in a resource-constrained IIoT. 

\subsubsection{Integrity}

Blockchain technology can address the challenges related to traceability and integrity, which are due to the complex supply chain management of today's world. However, blockchain does not provide a solution of the trust problem that is associated with the data in the IoT~\cite{malik2019trustchain}; and therefore, this problem leads to a research challenge of efficiently handling the trust problem with an optimization approach.

\subsubsection{Privacy and Security}

The IoT has unprecedented scalability and security challenges due to the high magnitude of data exchanged by devices in IoT platforms. By coupling blockchains with IoT applications, some of these challenges are addressed in~\cite{truong2019towards}. However, the design of a secure and scalable blockchain along with a resource optimization in the IoT is still a great challenge. 
The current Datagram Transport Layer Security (DTLS) and Transport Layer Security (TLS) are complex or heavy, especially for the IoT architecture with a massive number of devices~\cite{reyna2018blockchain}; and thus, there is a challenge to design proper security mechanisms as a part of a lower network layer while merging the IoT with blockchain.  

The primary concerns over the IoT platform are storage and secure data transaction. Blockchain can be an effective solution by utilizing its secure and decentralized structure. There are challenges of the implementation of blockchain over hardware-constrained devices~\cite{yakut2019blockchain}.
Since sensors are often used as IoT devices to detect the environment, a lot of sensitive data is collected in IoT platforms. Moreover, IoT devices can be highly exploited by attackers who maliciously use the IoT paradigm to launch various severe attacks~\cite{wu2019comprehensive}.
Therefore, it is required to design effective secure communication protocols, attack-defense mechanisms, and secure authentication techniques when integrating the IoT with blockchain.

\subsubsection{Management of Devices}

IoT architecture can have numerous sensors, and the management of the devices which are connected to the sensors remains a technical challenge forever. This challenge increases further while blockchain technology is merged with the IoT~\cite{yakut2019blockchain}. Therefore, an efficient design architecture is required while merging blockchain and the IoT in order to manage a huge number of sensors and devices connected in the IoT. This design can significantly help the mobile communication and satellite monitoring.

\subsubsection{Scalability}

Fog computing which is an extension of IoT-oriented solutions has the requirements for decentralization and distribution. In this regard, the integration of blockchain with IoT can provide a solution for decentralization and distribution, as well as can help fog computing to overcome the deficiencies in security and privacy~\cite{lei2020groupchain}. Therefore, an efficient design architecture integrating fog computing and blockchain is required to utilize the power of fog computing based IoT solutions. Scalability is one of the primary challenges of merging blockchain with fog computing~\cite{lei2020groupchain}.

\subsubsection{Consensus Mechanism}

IoT applications have high maintenance costs and weakness to support time-critical usages. These problems can be solved by proposing an appropriate distributed consensus mechanism~\cite{wu2019comprehensive}. The IoT platform has a huge number of various types of devices, and the communication cost becomes extremely high when every node in blockchain needs to broadcast its data and status. Therefore, it is required to overcome these obstacles and develop an efficient consensus mechanism to increase the overall throughput in the network while blockchain is combined with the IoT.

\subsubsection{Transaction Fees and Time}

One challenging issue in merging blockchain with the IoT implementation is the fees associated with the transactions in blockchain. These fees are created to reward miners but unfortunately, many cryptocurrencies allow to implement market-based fees, and therefore, the fees can become expensive~\cite{ensor2018blockchains}. Another problematic issue for blockchain is transaction time~\cite{ensor2018blockchains}. Bitcoin has a stable time span of ten minutes to create new  blocks. However, the transaction time is highly dependent on the congestion levels in the network~\cite{beck2016blockchain}. Hence, there is a large scope of research in the blockchain-based IoT payment system consisting of many micro-transactions which need to be finished within seconds.

\subsubsection{Transaction Throughput}

The heterogeneity of devices in the IoT platform implies different communication protocols, like Bluetooth, Wi-Fi, etc., and due to this heterogeneity, the cost of providing consensus in blockchain becomes expensive~\cite{wu2019comprehensive}. It is also non-trivial to impose a synchronization among the huge number of devices in the IoT. Consequently, improving the transaction throughput becomes challenging. Therefore, it is required to design proper communication protocols and good consensus techniques to improve the transaction throughput while the IoT and blockchain are combined together. The communication protocols should provide effective synchronization among various devices in the IoT.


\subsection{Challenges in Application of AI in IoT}

To develop an efficient intelligent IoT system, we need to identify the research challenges in the direction of integrating AI and the IoT together. Therefore, different challenges in applying AI in the field of the IoT are presented next.

\subsubsection{Training of AI-based Model}

The appropriate training of a ML model is required to make the system intelligent such that the system can take the proper decision dynamically based on the demand~\cite{son2020partial}. In this regard, the appropriate dataset needs to be applied and the dataset should be large enough to facilitate the learning process, by providing a sufficient information about the environment. In an IoT framework, different types of devices are deployed and various ML algorithms can be used to serve different purposes. Thus, it is a challenge to generate the appropriate and sufficient volume of datasets to efficiently train different ML algorithms, such that the whole system can adapt to the environment in order to provide several functionalities in the IoT. The performance of an intelligent system highly depends on the training procedure that is carried out to train the corresponding ML model. Therefore, the challenge is to perform the training procedure effectively and efficiently. Moreover, in distributed networks, the training of ML-based mechanisms is far challenging~\cite{chai2020hierarchical}.  

\subsubsection{Determination of Convergence}

A primary challenge in exploiting intelligence in a model is the proper analysis of convergence of the learning of ML-based models~\cite{tao2018primal}. The question of convergence is mainly raised in reinforcement learning where the system gains knowledge dynamically as the number of runs of the model increases~\cite{guo2019multi}. In this case, the convergence time influences the nature of adaptability of the ML module in different network scenarios~\cite{tao2019strength}. As the convergence time decreases, the ML module takes less time to gain knowledge about the environment. Therefore, our target is to minimize the convergence time of the learning of an intelligent system. Moreover, the convergence with the appropriate learning (knowledge) about the network is also important. Since an IoT model consists of a large number of varieties of devices, it is crucial to take care of the convergence of ML algorithms which are applied in the IoT framework~\cite{hussain2020new,tao2019strength}.

\subsubsection{Availability of Data for Training}

Data plays the primary role to train any ML-based model~\cite{ahamed2018applying}. To implement an AI-based IoT framework in a real case scenario, it is required to collect the data from client sides or through the actual positioning of sensors in the IoT. In this case, the designers of ML algorithms need to interact with domain experts to get the data. In addition, the collected data should have enough large volume so that the proposed ML model can be well-trained to solve the purposes of using this model in an IoT architecture. Therefore, the collection of the appropriate and sufficient large volume of data to train a ML model is a great challenge.

\subsubsection{Storage Management}

An AI-based system needs to store a huge volume of information such that, based on it, the system can take decision dynamically, and thus intelligence can be imposed in the system. Such high volume of information is required because it is difficult for a ML-based system to take decision on which information will be needed in future; and thus, the system stores all the data and information, which are obtained during the training of the model~\cite{kinkiri2016reducing}. Therefore, it is a challenge to efficiently handle the storage management while AI is used to make a system intelligent. In this case, the IoT yields a huge scope of research since a large number of varieties of devices are used in the IoT, where the edge-based implementation further provides a great challenge due to the engagement of end-users having low storage capabilities. 

\subsubsection{Development of Smart Cities} 

The idea of development of smart cities using proper biometric and AI recognition technologies can be developed further, along with an explainable AI for supporting high level human decisions~\cite{mochizuki2019ai}. The biometric solutions are used for police investigations, law enforcement and access to different facilities which will promote the safety in a much efficient way. Examples of such concepts have been tested by various countries and it has been found that, a proper exploitation of AI in developing smart cities will be really beneficial. For example, the city of Tigre in Argentina uses a command control centre with biometrics and video analytics, which led to the drop of city's car theft by $80\%$~\cite{mochizuki2019ai}. 
The AI is expected to give a decision for making a certain decision, and thus AI emerges the explainable AI field in the application of the IoT. This type of innovation can be achieved by focusing on the safety, security, and efficiency, which can be achieved by properly by using AI technologies in high levels of the IoT data exchange~\cite{mochizuki2019ai}. In addition, AI can be exploited in developing smart cities with the help of an edge computing architecture which can enhance the network performance and ease the development of a distributed architecture at client sides~\cite{bellavista2019support}.

\subsubsection{Cooking Innovations with AI}

Using IoT devices and special deep learning application, food calorie, category estimation, and training of robots for kitchen tasks are still a very underrated topic and should be explored further. By using IoT and AI technologies, it is possible to create a smart kitchen which can be monitored via mobile devices; and it will reduce the amount of physical labour that is put into the kitchen~\cite{francis2020automation}.

\subsubsection{IoT Device Security Based on Machine Learning} 

The IoT integrates devices in a network so that they can be used for executing different types of applications. It is really important to protect such devices from hackers such that the user privacy is maintained. Attacks like phishing, DoS, eavesdropping, and jamming can be prevented by using a variety of supervised and unsupervised learning algorithms which fall under artificial intelligence~\cite{xiao2018iot}. Moreover, we can use exposure indicators and predictive analytics for threats detection in such IoT networks. 
An efficient threat exposure and evaluation framework needs to be designed in order to improve preventive capabilities in the IoT, by combining AI-based tools for the continuous evaluation of security functionalities~\cite{brignoli2020combining}.

\subsubsection{Development of Integrated Circuit (IC) Technology with AI}

The ICs used for the IoT integrating AI need massive capabilities and smart integration. The ICs should be capable of concurrent engineering and multi-level co-optimization such that it can achieve the high performance, good yield, and low cost~\cite{chi2018fast}.

\subsubsection{Human Centric AI} 

The IoT uses huge data points and applies artificial intelligence to learn from individual user's actions and then use the knowledge gained from the learning to take appropriate decisions. Most of the AI models are purely based on mathematics, and this makes the AI model a black box, where users trust their personal experiences. This type of AI-based system learns from the inputs provided by users, and continuously improves with each new input. In the integration of artificial intelligence and IoT devices, the goal is to understand human emotion, human language, and behaviour; and thus, we can bridge the gap between machines and humans~\cite{garcia2019human}.

\subsubsection{AI-Enabled IoT for 5G Network Complexity}

The convergence of AI-enabled IoT, the network complexity analysis, and 5G allow for designing, prototyping, and enabling new systems that can adapt to various network scenarios. IoT systems in the 5G provide support for the applications in transportation, public safety, health, energy, and food productions. As the number of IoT devices connected to the Internet is increasing, the scalability and computability of these devices are becoming a challenge which needs approaches related to the network complexity analysis and designing new computational models~\cite{alvarado2019ai}.

\subsubsection{Wireless Networks}

The IoT requires continuous connectivity, dynamic service mobility, and extreme security when connected in a wireless network~\cite{ahmad2020challenges}. Artificial intelligence can play an important role in the underlying network infrastructure. In this case, the main challenges are designing of -- (i) AI-based approaches for handling bandwidth, spectrum, and latency constraints in wireless networks, (ii) AI-based security or caching requirement models with Big data, and (iii) an efficient architecture coping up with the network complexity in wireless environments. The generic needs of a wireless network specific design in the IoT can be implemented using the help of AI.

\subsubsection{Edge Computing on IoT Data Using AI}

IoT devices are continuously used and a huge amount of IoT data is generated. In order to extract insights from such high volume of data, the processing of data must happen in the place where it was generated~\cite{calo2017edge}. This type of process is called edge computing and it is a challenge as the AI algorithms need a lot of processing power which may not be available in the edge devices like laptops, mobile phones, etc. One particular way of architecture that can be used is a centralized approach-based AI model. 
In this way, it can be a possible option to reduce the overhead of executing AI applications in edge nodes. However, the next challenge will be to minimize the communication overhead between the central node and other edge devices.

\section{Conclusion}
\label{sec:conc}

Blockchain is a distributed immutable ledger system that can be used for implementing cryptocurrency with security; and AI is used to make an intelligent system. Therefore, these two mechanisms can strengthen functionalities of an IoT system. This survey presents an extensive study of the existing research works which have applied blockchain and AI in designing IoT-based applications. We first discuss about the IoT, blockchain, and AI, such that their potentialities can be understood while studying the impacts of applying blockchain and AI in IoT paradigms. We present a detailed discussion on the impacts of blockchain and AI in IoT-based applications which have been proposed in several existing works. In this direction, we also highlight the limitations of existing literature which have utilized blockchain and AI in the IoT. This survey also identifies the research challenges to integrate blockchain and AI with the IoT. These challenges open a fertile area for the future research scopes to exploit the capabilities of blockchain and AI in building IoT-based applications; and meaningful explorations of such challenges can lead to the development of smart IoT applications ensuring security, privacy, and trustworthy.



\bibliography{ref}


\end{document}